%% file: BLPS.tex
\documentclass[11pt]{article}
\pdfoutput=1
\usepackage[ngerman,english]{babel}
\usepackage[utf8]{inputenc}
\usepackage[T1]{fontenc}
\usepackage{lmodern}
\usepackage{float}
\usepackage{amssymb}
\usepackage{amsmath}
\usepackage{amscd}
\usepackage{bbm}
\usepackage{mathtools}
\usepackage{latexsym}
\usepackage{color}
\usepackage{cite}
\usepackage{tabularx}

\usepackage{hhline}
\usepackage[breaklinks,colorlinks,linkcolor=black,citecolor=black,
urlcolor=red]{hyperref}

\usepackage{tikz}

\catcode`@=11
\renewcommand{\section}{\@startsection{section}{1}{0pt}{\medskipamount}
{\medskipamount}{\large\bf}}
\numberwithin{equation}{section}
\catcode`@=12

\allowdisplaybreaks

\input{definitions}

\begin{document}
\begin{titlepage}
\setcounter{page}{0}
\begin{flushright}
ITP--UH--13/14
\end{flushright}

\vskip 1.5cm

\begin{center}

{\Large\bf Instantons on conical half-flat 6-manifolds
}

\vspace{12mm}
{\large Severin Bunk, Olaf Lechtenfeld${}^{\times}$, Alexander D. Popov
and Marcus Sperling}\\[8mm]

\noindent {\em
Institut für Theoretische Physik\\
Leibniz Universität Hannover \\
Appelstraße 2, 30167 Hannover, Germany }\smallskip\\
{Email: Olaf.Lechtenfeld, Alexander.Popov, Marcus.Sperling@itp.uni-hannover.de,\\ sb11@hw.ac.uk}\\[6mm]

\noindent ${}^{\times}${\em
Riemann Center for Geometry and Physics\\
Leibniz Universität Hannover \\
Appelstraße 2, 30167 Hannover, Germany }\\[6mm]

\vspace{12mm}

\begin{abstract}
\input{abstract}
\end{abstract}
\end{center}

\end{titlepage}
\tableofcontents
\vspace{0.3cm}
\hrule
\vspace{0.3cm}
\input{introduction}
\input{5-manifolds_and_SU2}
\input{Cones_and_SineCones}
\input{instantons_6-manifolds}
\input{conclusion}
\bibliographystyle{JHEP}     
\footnotesize{\bibliography{references}} 
\end{document}

%% file: definitions.tex
\renewcommand{\Im}{\mathrm{Im}}

\def\a{\alpha}
\def\b{\beta}
\def\g{\gamma}

\def\de{\delta}

\def\h{\eta}

\def\Th{\Theta}

\def\s{\sigma}

\def\o{\omega}
\def\Om{\Omega}
\def\La{\Lambda}

\def\1{\bar 1}
\def\2{\bar 2}
\def\3{\bar 3}
\def\4{\bar 4}

\newcommand{\Acal}{{\cal A}}

\newcommand{\with}{{\quad{\rm with}\quad}}

\newcommand{\ddu}[4]{{#1}_{#2#3}^{\phantom{#2#3}#4}}

\def\im{\mbox{i}}
\def\N2{$N{=}2$}

\def\diff{\mbox{d}}
\def\tr{{\rm tr}}
\def\sfrac#1#2{{\textstyle\frac{#1}{#2}}}
\def\>{\rangle}
\def\<{\langle}
\def\+{\dagger}
\def\={\ =\ }
\def\und{\qquad\textrm{and}\qquad}
\def\and{\quad\textrm{and}\quad}
\def\for{\quad\textrm{for}\quad}

%% file: abstract.tex
\noindent
We present a general procedure to construct $6$-dimensional manifolds with SU$(3)$-structure from SU$(2)$-structure $5$-manifolds.
We thereby obtain half-flat cylinders and sine-cones over $5$-manifolds with Sasaki-Einstein SU$(2)$-structure.
They are nearly Kähler in the special case of sine-cones over Sasaki-Einstein $5$-manifolds.
Both half-flat and nearly Kähler $6$-manifolds are prominent in flux compactifications of string theory.
Subsequently, we investigate instanton equations for connections on vector bundles over these half-flat manifolds.
A suitable ansatz for gauge fields on these $6$-manifolds reduces the instanton equation to a set of matrix equations.
We finally present some of its solutions and discuss the instanton configurations obtained this way.

%% file: introduction.tex
\section{Introduction}
\noindent
Solitons and instantons are important objects in modern field theory~\cite{Rajaraman:1982is,Manton:2004tk,Weinberg:2012}.
Solitons in supergravity theories are branes of various dimensions, which describe non-perturbative states of the underlying string theories or M-theory~\cite{Polchinski:Vol1,Polchinski:Vol2,Becker:2007}.
Branes in turn are sources of $p$-form flux fields.
They can also wrap various supersymmetric cycles of special holonomy 
manifolds~\cite{Becker:2007}, and these cycles (which are calibrated 
submanifolds \cite{Harvey:1982}) are defined, or calibrated, via the $p$-form 
fluxes.
Thus, fluxes play an important role in the compactification of low-energy string theories and M-theory.

String vacua with $p$-form fields along the extra dimensions (``flux 
compactifications'') have been intensively studied in recent years 
(see~\cite{Grana:2005jc,Blumenhagen:2006} for a review and references).
In particular, fluxes in heterotic string theories, which play a prominent role in string-theory model building due to the easy incorporation of the standard-model gauge group, have been considered
e.g.\ in~\cite{Louis:2001uy,Louis:2006kb,Cardoso:2003af,Frey:2005zz,Manousselis:2005xa,
Becker:2003yv,Becker:2003sh,Fernandez:2008wa,Papadopoulos:2009br,Held:2010az,Lechtenfeld:2010dr,
Klaput:2011mz,Harland:2011zs,Gemmer:2012pp}.
Heterotic flux compactifications have been known for quite some time, starting from refs.~\cite{Hull:1986iu,Hull:198651,Hull:1986357,Strominger:1986uh} in the mid-1980s.
On Calabi-Yau manifolds the introduction of fluxes partially resolves the vacuum 
degeneracy problem by giving masses to problematic moduli, but they lead to 
non-integrable SU$(3)$-structures (i.e. with intrinsic torsion) on the internal 
compact $6$-manifolds.
Among these manifolds there are six-dimensional nearly Kähler and half-flat manifolds \cite{Louis:2001uy,Louis:2006kb,Cardoso:2003af,Frey:2005zz,Manousselis:2005xa,
Harland:2011zs,Gemmer:2012pp,Klaput:2011mz,Lechtenfeld:2010dr}.

Heterotic supergravity, as a low-energy effective field theory, preserves 
supersymmetry in $10$ dimensions precisely if there exists at least one 
globally defined Majorana-Weyl spinor $\epsilon$ such that the supersymmetry 
variations of the fermionic fields (gravitino $\lambda $, dilatino $\phi$, and 
gaugino $\xi$) vanish, i.e. the so-called BPS equations
\begin{subequations}
\label{eqn:BPS}
\begin{align}
 \delta_{\lambda} &= \nabla^+ \epsilon =0 \; , \label{eqn:gravitino} \\
\delta_{\psi} &= \gamma\left( \diff \phi - \frac{1}{2} H \right) \epsilon =0 \; 
, 
\label{eqn:dilatino} \\
\delta_{\xi} &= \gamma \left(\mathcal{F}_{\mathcal{A}} \right) \epsilon=0 
\label{eqn:gaugino}
\end{align}
\end{subequations}
hold, wherein $\gamma(\omega) = \frac{1}{p!} \omega_{i_1 \ldots  i_p} 
\Gamma^{i_1 \ldots  i_p}$ is the Clifford map for a $p$-form $\omega$. The 
bosonic field content is given by the metric $g$, the dilaton $\phi$, the 
$3$-form $H$, and the gauge field $\mathcal{A}$. Further, $\nabla^+$ is a 
metric compatible connection with 
torsion $H$. 

The $10$-dimensional space is assumed to be a product $\mathbb{M}^{p-1,1} 
\times M^{10-p}$, where $M^{10-p}$ is a $d=(10-p)$-dimensional internal 
manifold. 
Then~\eqref{eqn:gravitino} translates into the existence of an covariantly 
constant spinor $\epsilon_d$ on $M^d$.  
Moreover, a globally defined non-vanishing spinor exists only on manifolds 
$M^d$ with reduced structure group (i.e. a $G$-structure), which in $d=6$ 
amounts to an SU$(3)$-structure. Then a metric compatible connection, which 
leaves $\epsilon_6$ parallel and is also compatible with the SU$(3)$-structure, 
always exists, but possibly has torsion. In other words, a connection with 
SU$(3)$-holonomy always exists on SU$(3)$-manifolds. As a consequence, 
manifolds with special holonomy or $G$-structure are essential in string 
theory compactifications. Moreover, $G$-structures then allow for a 
$(d-4)$-form $\Psi$ on $M^d$, such that the natural choice for the $3$-form flux 
is $H=*\diff \Psi$. 

In addition, the curvature $\mathcal{F}_{\mathcal{A}} $ of a connection 
$\mathcal{A}$ on a gauge bundle has to satisfy the generalized instanton 
equation~\eqref{eqn:gaugino}.
In particular, the instanton equation can be introduced on any 
manifold with a $G$-structure. On manifolds $M^d$ with integrable 
$G$-structure, instantons have two crucial features. First, they solve the 
Yang-Mills equation (without torsion), and, second, the Levi-Civita connection 
on $TM^d$ already is 
an instanton.
 
The BPS equations~\eqref{eqn:BPS} 
have to be supplemented by the $\alpha'$-corrected Bianchi identity 
\begin{equation}
 \diff H = \frac{\alpha'}{4} \left[ \tr\left( R \wedge R \right) - \tr\left( 
\mathcal{F}_{\mathcal{A}} \wedge \mathcal{F}_{\mathcal{A}}  \right) \right] 
\label{eqn:Bianci}
\end{equation}
due to the Green-Schwarz anomaly cancellation mechanism. Here $R$ is the 
curvature of a connection $\nabla$ on the tangent bundle\footnote{Different 
choices for $\nabla$, such as $\nabla^+$, are mentioned 
in~\cite{Ivanov:2009rh}.}. For compactifications with $\diff H \neq 0$ one has 
the additional 
freedom to choose a gauge bundle compared to Calabi-Yau compactifications, 
wherein the vanishing of $\diff H$ can be achieved by the ``standard 
embedding'' of the spin connection $\nabla^+$ into the gauge 
connection $\mathcal{A}$, i.e. the gauge bundle is just $TM^d$. However, the 
choice of the gauge bundle for $\diff H \neq 0$ is restricted by the 
Bianchi identity and the instanton equations (which on Kähler manifolds are 
just the Donaldson-Uhlenbeck-Yau 
equations~\cite{Donaldson01011985,donaldson1987,Uhlenbeck:CPA1986} that 
correspond to a stability criterion on holomorphic bundles).

By a theorem of Ivanov~\cite{Ivanov:2009rh}, a solution to the 
BPS equations~\eqref{eqn:BPS} and the Bianchi identity~\eqref{eqn:Bianci} 
satisfies the heterotic equations of motion if and only if the connection 
$\nabla$ is an SU$(3)$-instanton in $d=6$. In other words, $R$ and 
$\mathcal{F}_{\mathcal{A}}$ are treated on the same footing in a pure 
supergravity view, i.e. $\gamma(\mathcal{F}_{\mathcal{A}}) \epsilon 
=\gamma(R)\epsilon=0 $. Therefore, in the spirit 
of~\cite{Harland:2011zs,Gemmer:2012pp,Chatzistavrakidis:2012qb,Gemmer:2013ica}, 
we study the 
instanton equation~\eqref{eqn:gaugino} for non-integrable SU$(3)$-structures in 
order to provide an important ingredient for full heterotic supergravity 
solutions\footnote{Choosing a different connection $\nabla$, for example 
$\nabla^+$, the BPS equations together with the Bianchi identity imply the 
heterotic equations of motion only up to higher $\alpha'$-correction. This 
yields a perturbative solution in contrast to the exact solution advocated 
above.}.

The construction of metric cones and sine-cones over manifolds $M^d$ with a 
$G$-structure provides a tool to generate and link different $G'$-structures on 
$(d+1)$-dimensional manifolds.
Most prominently, Sasaki-Einstein $5$-manifolds generate a Calabi-Yau structure on their metric cone and a nearly Kähler structure on their sine-cone.
A generalization is possible by means of the notion of hypo geometry, in 
particular to hypo, nearly hypo and double hypo SU$(2)$-structures; see for 
instance~\cite{Fernandez:2006ux}.
Double hypo structures lift to nearly Kähler as well as to half-flat SU$(3)$-structures on the sine-cone.
The described ``linking'' phenomenon is well-known from the cases of cylinders, cones and sine-cones over nearly Kähler 6-manifolds, which lead to different G$_2$-manifolds~\cite{Gemmer:2011cp}.
Here we use these techniques in order to construct $6$-dimensional manifolds 
with special SU$(3)$-structures that may be valuable, for example, in flux 
compactifications of the heterotic string.

Supergravity in 10 dimensions allows for brane solutions which interpolate 
between an Ad$S_{p+1}\times M^{9-p}$ near-horizon geometry and an asymptotic
geometry $\mathbb{R}^{p-1,1}\times C(M^{9-p})$, where $C(M^{9-p})$ is a metric 
cone over $M^{9-p}$ (see e.g.~\cite{Acharya:1999,Koerber:2008rx}
and references therein). These brane solution in heterotic 
supergravity with Yang-Mills instantons on the metric cones $C(M^{9-p})$ have 
been considered in~\cite{Harland:2011zs,Gemmer:2012pp,Nolle:2012hf}.
Here, we take the first step to generalize them by considering sine-cones with 
nearly Kähler structures as well as cylinders with half-flat structures instead 
of metric cones with Kähler structures.

The generalization of Yang-Mills instantons to higher dimensions ($d{>}4$) was
first proposed in~\cite{Corrigan:1982th} and further studied
in~\cite{Donaldson01011985,donaldson1987,Uhlenbeck:CPA1986,
Capria:1988,Carrion:1998,BLKHS:1998,Blau:1997,Acharya:1998,Donaldson:1996,
Donaldson:2009,Tian:2000} (see also references therein). Some solutions for
$d{>}4$ have been found, namely $\mathrm{Spin}(7)$-instantons on $\mathbb{R}^8$
in~\cite{Fairlie:1984mp,Fubini:1985jm} and G$_2$-instantons on $\mathbb{R}^7$
in~\cite{Ivanova:1992nj,Ivanova:1993ws,Gunaydin:1995ku}.
For generic non-integrable $G$-structures, the instanton equation
implies the Yang-Mills equation with torsion. However, as shown
in~\cite{Harland:2011zs}, on manifolds with real Killing spinors the corresponding instantons solve the Yang-Mills equation without torsion even if the $G$-structure has non-vanishing intrinsic torsion.
Recently, we constructed instantons on Kähler-torsion and hyper-Kähler-torsion sine-cones over Sasakian manifolds in~\cite{Bunk:2014kva}.
In this paper we extend these studies to nearly Kähler sine-cones and half-flat cylinders over Sasaki-Einstein manifolds.

The outline of the paper is as follows: Section~\ref{sec:SU(2)-structures} is 
devoted to a review of various SU$(2)$-structures, focusing on
hypo geometry and investigating the $5$-sphere as an example. 
Section~\ref{sec:Cones_and_SineCones} then
provides several cone constructions that link Sasaki-Einstein $5$-manifolds to particular SU$(3)$ $6$-manifolds.
In Section~\ref{sec:Instantons} the instanton equations on these $6$-dimensional 
SU$(3)$-manifolds are derived, and utilizing a certain ansatz for the gauge 
connection these equations are reduced to matrix equations. We derive some 
particular solutions for these matrix equations by the choice of a 
suitable matrix ansatz and discuss their corresponding gauge field 
configurations. 

%% file: 5-manifolds_and_SU2.tex
\section{SU(2)- and SU(3)-structures in 5 and 6 dimensions}
\label{sec:SU(2)-structures}
\subsection{Sasakian structures}
We begin by introducing several geometric structures that will become important 
in the constructions of this paper.
As in~\cite{Sparks:2010}, an \emph{almost contact metric manifold} is an
odd-dimensional Riemannian
manifold $(M^{2m+1},g)$ such that there exists a reduction of the structure
group $SO(2m+1)$ of the bundle of orthonormal frames on $TM$ to U$(m)$. For such
manifolds there exist a 1-form $\h$ and a 2-form $\o$ such that $\h \wedge
(\o)^m \neq 0$. \emph{Contact metric structures} are characterized by $\diff \h =
2\, \o$ in our sign convention.

An almost contact structure is characterized by the Nijenhuis torsion tensor~\cite{Boyer:2008} $\mathfrak{N}=(\mathfrak{N}_{\mu \nu}^\sigma)$.
A \emph{quasi-Sasakian structure} is given by $\mathfrak{N}=0$ and $\diff \o =0$.
In particular, if $\diff \h = \a\, \o$ with $\a \in \mathbb{R}$, then the almost
contact structure is called
\emph{$\a$-Sasakian}. If $\a = 2$ the structure is called \emph{Sasakian}.

Let us now specialize to the 5-dimensional case $M^5$, and let $e^\mu$ be an 
orthonormal coframe with $\mu = (a,5)$ and $a = 1,\ldots,4$. Sasakian 
$5$-manifolds are endowed with a 1-form $\h$, 2-form $\o$, 3-form $P$, and 
4-form $Q$ satisfying the relations
\begin{subequations}
 \begin{alignat}{2}
  \eta &= -e^5\;, & \quad \o &\equiv \o^3 = \sfrac{1}{2}\, \h_{ab}^{3}\, e^a \wedge e^b\;,
  \qquad  \h \lrcorner\, \o^3 = 0\;,\\
  P &= \o^3 \wedge \h\;, & \quad Q &= \sfrac{1}{2}\, \o^3 \wedge \o^3\;,\\
  \diff \h &= 2\, \o^3\;, & \quad \diff P &= 4\, Q\;.
 \end{alignat}%
\end{subequations}%
Here $\h^{3}_{ab}$ are the components of the self-dual
't~Hooft tensors~\cite{Rajaraman:1982is}, and the contraction of two forms is defined as $\eta \lrcorner\, \omega \coloneqq \ast(\eta \wedge \ast \omega)$ (see e.g.~\cite{Harland:2009yu}).
\subsection{SU(2)-structures in $d = 5$}
\label{subsec:SU2-structure}
Let $M^5$ be $5$-manifold with an SU$(2)$-structure, i.e.~the frame bundle of
$M^5$ can be reduced to an SU$(2)$ principal subbundle. It has been proven in
\cite{Conti:2005} that an SU$(2)$-structure is determined by a
quadruplet $(\h, \o^{1} , \o^{2} , \o^{3})$ of differential forms, wherein
$\eta \in \Om^1(M^5)$ and $\o^\a \in \Om^2(M^5)$ for $\a = 1,2,3$. These forms
satisfy 
\begin{align}
 \o^\a \wedge \o^\b =2\, \de^{\a \b} Q
\end{align}
for the 4-form $Q = \frac{1}{2}\, \omega^3 \wedge \omega^3$ with $\h \wedge Q \neq 0$.

Moreover, it has been shown in \cite{Conti:2005} that it is always possible to choose a
local orthonormal basis $e^1, \ldots, e^5$ of forms on $M^5$ such that
\begin{align}
 \h = - e^5 \; ,  \qquad 
 \o^1 = e^{23} + e^{14} \;, \qquad
 \o^2 = e^{31} + e^{24} \;, \qquad 
 \o^3 = e^{12} + e^{34} \;.
\label{eqn:local_SU(2)-forms}
\end{align}
By means of the 't~Hooft symbols $\h_{ab}^\a$, one can express the 2-forms as 
\begin{align}
 \o^\a = \sfrac{1}{2}\, \h_{ab}^\a\, e^a \wedge e^b. %
\label{eqn:tHooft_symbols}
\end{align}
Here again $a,b = 1,2,3,4$.
Among the SU$(2)$-structures in $5$ dimensions there are several types having particularly interesting geometry.
We will now recall their definitions following~\cite{Fernandez:2006ux}.

\paragraph{Sasaki-Einstein:}
A Sasaki-Einstein $5$-manifold is a manifold carrying an SU$(2)$-structure 
defined by $(\h, \o^{1} , \o^{2} , \o^{3})$, where these forms are subject to
\begin{align}
 \diff \, \h = 2\, \o^3 \; , \qquad \diff \,  \o^1 = - 3\, \h \wedge \o^2 \; , 
\qquad \diff \,
 \o^2 = 3\, \h \wedge \o^1 \; . %
\label{eqn:def_SE5}
\end{align}
\paragraph{Hypo:} An SU$(2)$-structure on a 5-manifold is called \emph{hypo} if
\begin{align}
 \diff \, \o^3 =0 \; , \qquad \diff {\left( \o^1 \wedge \h \right)} = 0 \; , 
\qquad \diff {\left( \o^2 \wedge \h \right)}=0 %
\label{eqn:def_hypo}
\end{align}
holds true. Hypo geometry, therefore, is a generalization of Sasaki-Einstein
geometry.
\paragraph{Nearly hypo:} An SU$(2)$-structure on a 5-manifold is called
\emph{nearly hypo} if it satisfies 
\begin{align}
 \diff \, \o^1 = - 3\, \h \wedge \o^2 \; , \qquad \diff \left( \h \wedge \o^3 
\right) = 2\, \o^1 \wedge \o^1 \; .%
\label{eqn:def_nearly_hypo}
\end{align}
Note that any SU$(2)$ structure which satisfies the first two
identities of \eqref{eqn:def_SE5} is a nearly hypo structure.
\paragraph{Double hypo:} An SU$(2)$-structure on a 5-manifold is called
\emph{double hypo} if it is hypo and nearly hypo simultaneously, i.e.~if it 
satisfies \eqref{eqn:def_hypo} and \eqref{eqn:def_nearly_hypo}. Thus, the 
Sasaki-Einstein
$5$-manifolds are a subset of the double hypo manifolds.\medskip\\
As shown in~\cite{Conti:2005}, SU$(2)$-structures in $5$ dimensions always induce a nowhere-vanishing spinor on $M^5$.
This will be generalized Killing if and only if the SU$(2)$-structure is hypo, and Killing if and only if the SU$(2)$-structure is Sasaki-Einstein.
In~\cite{Harland:2011zs} it has been argued that in the latter case there exists a one-parameter family of metrics
\begin{align}
 g_{M^5}=e^{2h} \de_{ab}\,e^a \otimes e^b + e^5 \otimes e^5 
\end{align}
which is compatible with an $\mathfrak{su}(2)$-valued connection on $M^5$ for which the Killing spinor is parallel.
For the special value $\exp(2h)=4/3$ the torsion of that connection is totally 
antisymmetric and parallel with respect to that connection, i.e.~there exists a 
canonical $\mathfrak{su}(2)$ connection.
For all values of $h$ however, this connection is an $\mathfrak{su}(2)$ instanton on $TM^5$  for the respective SU$(2)$-structure.
For $h=0$, $M^5$ is a Sasaki-Einstein manifold and the torsion components of 
the canonical connection read
\begin{align}
 T^a=\sfrac{3}{4} P_{a \mu \nu} e^{\mu \nu}  \und  T^5=P_{5 \mu \nu} e^{\mu\nu} \; .
\end{align}
\subsection{Example: the $5$-sphere}
We illustrate how different types of SU$(2)$-structures are embedded into each other with the example of the
$5$-sphere written as the homogeneous space $S^5=\mathrm{SU}(3)/\mathrm{SU}(2)$.

The SU$(3)$-structure constants can be chosen as 
\begin{subequations}
\begin{align}
  \ddu{f}{3}{1}{6}&= -\ddu{f}{2}{4}{6} = \ddu{f}{2}{3}{7} = -\ddu{f}{1}{4}{7}
= \ddu{f}{1}{2}{8}= -\ddu{f}{3}{4}{8}=\frac{1}{2\sqrt{3}} \; ,\\%
 \ddu{f}{6}{7}{8}&= \ddu{f}{8}{6}{7}= \ddu{f}{7}{8}{6}= \frac{1}{\sqrt{3}} \;,
\\%
\ddu{f}{1}{2}{5}&= \ddu{f}{3}{4}{5}= -\frac{1}{2} \,,%
\end{align}%
\label{eqn:structure_constants}%
\end{subequations}  %
by using rescaled Gell-Mann matrices as a basis of
$\mathfrak{su}(3)$. The structure constants~\eqref{eqn:structure_constants} are
completely antisymmetric upon permutation of indices, and all other index
combinations are zero.
The Cartan-Killing form is then given by
\begin{align}
  \ddu{f}{A}{D}{C} \ddu{f}{C}{B}{D} = \de_{AB}\;, \quad A,B,C,D \in
\{1,2,\ldots,8\}\, .%
\label{eqn:Cartan_Killing}
\end{align}
As local coframes $e^\mu=(e^a,e^5)$ on the coset space  we use the images of the left-invariant $1$-forms on
SU$(3)$ under a pull-back along a section of the SU$(2)$ principal bundle $\mathrm{SU}(3)
\to \mathrm{SU}(3)/\mathrm{SU}(2)$.
The coset with the structure constants~\eqref{eqn:structure_constants} is equipped with the Cartan-Killing metric, which can then be expressed as
($a,b,c,d=1,\ldots,4$ and $i=6,7,8$)
\begin{subequations}
\begin{alignat}{1}
 g_{ab} &= \ddu{f}{a}{d}{c} \ddu{f}{c}{b}{d} + 2 \ddu{f}{a}{5}{c} \ddu{f}{c}{b}{5} 
 + 2 \ddu{f}{a}{i}{c} \ddu{f}{c}{b}{i} = \delta_{ab} \; ,\\
g_{55} &=\ddu{f}{5}{d}{c} \ddu{f}{c}{5}{d} = \delta_{55} \; . 
\end{alignat}%
\label{eqn:round_metric}%
\end{subequations}
The use of left-invariant objects on SU$(3)$ enables us to explicitly compute 
connection components from the Maurer-Cartan equation.
The connection $1$-forms $\Gamma^\mu_\nu$ and the torsion $2$-forms $T^\mu$ are
then
given as
\begin{align}
 \diff e^\mu = -\ddu{f}{i}{\nu}{\mu} e^i \wedge e^\nu - \sfrac{1}{2}
\ddu{f}{\nu}{\sigma}{\mu} e^\nu \wedge e^\sigma = - \Gamma^\mu_\nu \wedge e^\nu
+ T^\mu\; ,
\end{align}
such that
\begin{align}
 T^\sigma = \sfrac{1}{2} T_{\mu \nu}^\sigma e^\mu \wedge e^\nu \quad \Rightarrow
\quad T_{\mu \nu}^\sigma = - \ddu{f}{\mu}{\nu}{\sigma} \; .
\end{align}
With the Cartan-Killing metric~\eqref{eqn:round_metric} one obtains the totally antisymmetric components
\begin{align}
 T_{a5b} = -f_{a5b} = f_{5ab} \; .
\end{align}
Note that
\begin{align}
 \diff e^5 = - \ddu{f}{a}{b}{5} e^a\wedge e^b = \sfrac{1}{2}\, \o^3\;,
\end{align}
and that $\mathrm{SU}(3)/\mathrm{SU}(2)$ is endowed with an SU$(2)$-structure given by $e^5$ and $\o^\a$
as defined in~\eqref{eqn:local_SU(2)-forms}.
As a canonical connection on $\mathrm{SU}(3)/\mathrm{SU}(2)$ we have with the above choices
\begin{align}
 {}^{(c)}\Gamma_b^a = {f_{ib}}^a\, e^i = {}^{(c)}\Gamma_{\mu b}^a\, e^\mu \; .
\end{align}

Now, we introduce a two-parameter family of SU$(2)$-structures on $S^5$ by a
rescaling of the $\mathfrak{su}(3)$ generators. Consider
\begin{align}
 I_a \to \tilde{I}_a = \frac{1}{\beta}\, I_a \; , \qquad I_5 \to \tilde{I}_5 =
\frac{1}{\gamma}\, I_5
\; , \qquad I_i \to \tilde{I}_i = I_i
\end{align}
for $(\gamma,\beta) \in (\mathbb{R}\setminus \lbrace 0 \rbrace) \times
\mathbb{R}^+$.
(A change of sign for $\beta$ does not define a different SU$(2)$-structure.)
Consequently, the structure constants are changed as follows,
\begin{subequations}
\begin{alignat}{2}
 \ddu{f}{5}{b}{a} &\to  \ddu{\tilde{f}}{5}{b}{a}  = \frac{1}{\gamma}\,
 \ddu{f}{5}{b}{a} \; ,& \qquad
 \ddu{f}{a}{b}{5} &\to \ddu{\tilde{f}}{a}{b}{5} =
\frac{\gamma}{\beta^2}\, \ddu{f}{a}{b}{5}  \; ,  \\
 \ddu{f}{a}{b}{i} &\to \ddu{\tilde{f}}{a}{b}{i} =
\frac{1}{\beta^2}\, \ddu{f}{a}{b}{i} \; , & \qquad \ddu{f}{i}{b}{a}  
&\to \ddu{\tilde{f}}{i}{b}{a} = \ddu{f}{i}{b}{a} \; , \\
 \ddu{f}{j}{k}{i} &\to \ddu{\tilde{f}}{j}{k}{i} = \ddu{f}{j}{k}{i} \; . & & 
\end{alignat}
\label{eqn:structure_constants_rescaled}%
\end{subequations}
A rescaling of the generators of $\mathfrak{su}(3)$ rescales the left-invariant
vector fields and $1$-forms accordingly, and this is propagated to the coset via
the pullback as used before.
In particular, the rescaled structure constants have to be used in the Maurer-Cartan equation in order to compute the differentials of the rescaled $\tilde{e}^\mu$.
We can use~\eqref{eqn:local_SU(2)-forms} with respect to the new coframes $\tilde{e}^\mu$ to define
a rescaled SU$(2)$-structure on $S^5$.
The differentials of the defining forms then read
\begin{subequations}
\label{eqn:SU2_family_on_S5}
 \begin{align}
  \diff \tilde{\h}  &= -\frac{\g}{2\b^2}\, \tilde{\o}^3\; ,\\%
  \diff\tilde{\o}^1 &= \frac{1}{\g}\, \tilde{\h} \wedge \tilde{\o}^2\; ,\\%
  \diff\tilde{\o}^2 &= -\frac{1}{\g}\, \tilde{\h} \wedge \tilde{\o}^1\; ,\\%
  \diff\tilde{\o}^3 &= 0\; .%
 \end{align}%
\end{subequations}%
Thus, $(\tilde{\eta},\tilde{\omega}^1,\tilde{\omega}^2,\tilde{\omega}^3)$ is a 
two-parameter family of hypo SU$(2)$-structures on $S^5$, as the 
conditions~\eqref{eqn:def_hypo} are
satisfied for all values of $\b$ and $\g$.
For the value $(\gamma,\beta) = (-\frac{1}{3}, \frac{1}{2\sqrt{3}})$ this turns
out to be nearly hypo additionally, and, as a consequence, at this value the
SU$(2)$-structure is double hypo.
Furthermore, this particular SU$(2)$-structure is even Sasaki-Einstein, as we also show by a direct calculation of the Ricci tensor below. Therefore, the family of SU$(2)$-structures on $S^5$ does not discriminate between the double hypo and Sasaki-Einstein property.
However, it shows how, by a simple rescaling of the generators of
$\mathfrak{su}(3)$, one can induce different SU$(2)$-structure geometries on
$S^5$.

Note that there are many possible choices of a Riemannian metric on the coset space.
Among them are the Cartan-Killing metric and the round metric on $S^5$, which 
we consider in the following:
\paragraph{Cartan-Killing metric:}
From the definition~\eqref{eqn:Cartan_Killing} we obtain
\begin{equation}
 g_{\mathrm{CK}} = \delta_{ab}\, e^a \otimes e^b + e^5 \otimes e^5 \;.
\end{equation}
We express this with respect to local frames $\tilde{e}$ adapted to the Sasaki-Einstein SU$(2)$-structure (i.e.~for $(\gamma,\beta) = (-\frac{1}{3}, \frac{1}{2\sqrt{3}})$).
Thus, we arrive at
\begin{equation}
 g_{\mathrm{CK}}= 12 \, \delta_{ab}\, \tilde{e}^a \otimes \tilde{e}^b + 9 \, \tilde{e}^5 \otimes \tilde{e}^5\; .
\end{equation}
By means of the Maurer-Cartan equations
\begin{subequations}
\begin{align}
 \diff \tilde{e}^\mu &= -\sfrac{1}{2}\, \ddu{\tilde{f}}{\nu}{\rho}{\mu}\, \tilde{e}^{\nu} \wedge \tilde{e}^{\rho} - \ddu{\tilde{f}}{i}{\nu}{\mu}\, \tilde{e}^i \wedge \tilde{e}^{\nu} \; , \\ %
 \diff \tilde{e}^i &= -\sfrac{1}{2}\, \ddu{\tilde{f}}{j}{k}{i}\, \tilde{e}^j \wedge \tilde{e}^k - \sfrac{1}{2} \ddu{\tilde{f}}{\mu}{\nu}{i}\, \tilde{e}^{\mu} \wedge \tilde{e}^{\nu}
\end{align}
\end{subequations}
and demanding the torsion $2$-form $T^\mu$ to vanish, one obtains
\begin{align}
 {}^{\mathrm{CK}}\Gamma_{b}^{a} = \ddu{\tilde{f}}{i}{b}{a}\, \tilde{e}^i + \sfrac{1}{2}\, \ddu{\tilde{f}}{c}{b}{a}\, \tilde{e}^c
\end{align}
for the connection $1$-forms of the Levi-Civita connection induced by the Cartan-Killing metric on $S^5 = \mathrm{SU}(3)/\mathrm{SU}(2)$.
The curvature $2$-form 
\begin{align}
{}^{\mathrm{CK}} R_b^a = \diff \, {}^{\mathrm{CK}}\Gamma_{b}^{a} + 
{}^{\mathrm{CK}}\Gamma_{c}^{a} \wedge  {}^{\mathrm{CK}}\Gamma_{b}^{c}
\end{align}
can be computed, and all $2$-form contributions proportional to $\tilde{e}^j
\wedge \tilde{e}^k $ or $ \tilde{e}^j \wedge \tilde{e}^\mu$ vanish due to
the Jacobi identity~\cite{MuellerHoissen:1987}. Thus, the Ricci tensor reads
\begin{align}
 {}^{\mathrm{CK}}\mathrm{Ric}_{ab} &= \ddu{\tilde{f}}{a}{i}{c} \ddu{\tilde{f}}{c}{b}{i} + \sfrac{1}{4} \left( \ddu{\tilde{f}}{a}{c}{5} \ddu{\tilde{f}}{5}{b}{c} + \ddu{\tilde{f}}{a}{5}{c} \ddu{\tilde{f}}{c}{b}{5} \right)
 = \sfrac{9}{2}\, \delta_{ab}\;,\\%
 {}^{\mathrm{CK}}\mathrm{Ric}_{55} &= \sfrac{1}{4}\, \ddu{\tilde{f}}{5}{d}{c} \ddu{\tilde{f}}{c}{5}{d} = \sfrac{9}{4} \; , \qquad  {}^{\mathrm{CK}}\mathrm{Ric}_{a5} =0\; .
\end{align}
This shows that the choice of structure constants~\eqref{eqn:structure_constants} yields an $\a$-Sasakian manifold with $\a = -\frac{1}{2}$ (c.f. equation~\eqref{eqn:SU2_family_on_S5} for $\gamma = \beta = 1$), but not an Einstein space.
\paragraph{Round metric:} Using again the local coframes $\tilde{e}$ adapted to the Sasaki-Einstein structure, the metric induced by stereographic projection 
from
the ambient $\mathbb{R}^6$ reads 
\begin{align}
 g_{\mathrm{rnd}}=  \delta_{ab}\, \tilde{e}^a \otimes \tilde{e}^b + \tilde{e}^5 \otimes \tilde{e}^5
 = \delta_{\mu\nu}\, \tilde{e}^\mu \otimes \tilde{e}^\nu \; .
\end{align}
Employing the Koszul formula for the round metric and the coframes
$\tilde{e}^\mu$, one can calculate the Christoffel symbols of the Levi-Civita
connection to be
\begin{equation}
 {}^{\mathrm{rnd}}\Gamma_{\mu \nu}^{\rho} = \sfrac{1}{2}\, \ddu{\tilde{f}}{\mu}{\nu}{\rho} - 2\, \tilde{f}_{(\mu \ \nu)}^{\ \ \rho}\; .
\end{equation}
As before, the computation of
the Ricci tensor is straightforward, and the result for this case is
\begin{align}
{}^{\mathrm{rnd}} \mathrm{Ric}_{\mu \nu} = 4 \, (g_{\mathrm{rnd}})_{\mu \nu} = 4\, \delta_{\mu \nu} \;
.
\end{align}
Hence, the $5$-sphere endowed with the round metric is an Einstein space with Einstein constant $4$, just as expected.
\subsection{SU(3)-structures in $d = 6$}
\label{sec:SU3_structures}
As pointed out in the introduction, one of our goals is the construction of 
SU$(3)$-structures on $6$-dimensional manifolds.
Therefore, we introduce these structures and their characterization via intrinsic torsion classes.
In a manner similar to Subsection~\ref{subsec:SU2-structure}, an
SU$(3)$-structure on a 6-manifold $M^6$ is given by a reduction of the
frame bundle to an SU$(3)$ subbundle. An SU$(3)$-structure on a $6$-dimensional
manifold $M^6$ is characterized in terms of a triple $(J,\o,\Om)$, where $J$ is an
almost complex structure, $\o$ a $(1,1)$-form, and $\Om$ a
$(3,0)$-form with respect to $J$. These are subject to the algebraic relations
\begin{subequations}
 \begin{align}
  \o \wedge \Om ={}& 0\; ,\\
  \Om \wedge \bar{\Om} ={}& - \sfrac{4\im}{3}\, \o \wedge \o \wedge \o\; .
 \end{align}
\end{subequations}
The compatible Riemannian metric is determined by 
$\omega(\cdot,\cdot) = g(J(\cdot),\cdot)$, and the $(3,0)$-form can be split 
into its real and imaginary part, i.e.~$\Omega =
\Omega^+ + \im\, \Omega^-$.
By an appropriate choice of a local frame, these forms can always be 
brought into the form
\begin{equation}
\label{eqn:SU3_standard_components}
 \omega = e^1 \wedge e^2 + e^3 \wedge e^4 + e^5 \wedge e^6 \and
 \Omega = (e^1 + \im e^2) \wedge (e^3 + \im e^4) \wedge (e^5 + \im e^6).
\end{equation}

For SU$(3)$-structures in $6$ dimensions, there exist several types of such 
structures with different geometric behavior, which is mostly governed by the 
differentials $\diff \omega$ and $\diff \Omega$.
SU$(3)$-structures in $6$ dimensions have been classified in terms of their five
intrinsic torsion classes~\cite{Chiossi:2002tw}.
These are encoded in the differential of the defining forms in the following
manner:
\begin{subequations}
\begin{align}
\diff \o &= \sfrac{3}{2}\, \Im\left( \left( W_1^+ - \im W_1^- \right) \Om \right) +
W_3 + W_4 \wedge \o \; , \\
\diff \Om &= \left( W_1^+ + \im W_1^- \right) \o \wedge \o  + \left( W_2^+ + \im
W_2^- \right) \wedge \o + \Om \wedge W_5 \; . 
\end{align}
Here $W_1^\pm$ are real functions, $W_4$ and $W_5$ are real 1-forms, $W_2^\pm$ are the real and imaginary part of a $(1,1)$-form, respectively, and $W_3$ is the real part of a $(2,1)$-form.
Note that both $W_2$ and $W_3$ are primitive forms~\cite{Grana:2005jc}, i.e.
\begin{equation}
 \omega \lrcorner W_2 = 0 \and \omega \lrcorner W_3 = 0.
\end{equation}
\end{subequations}

The Nijenhuis tensor gives rise to the components $W_1$ and $W_2$; thus, the almost complex structure $J$ of any SU$(3)$-structure with non-vanishing 
$W_1$ or $W_2$ is non-integrable.

To finish this section, let us list the structures of particular relevance 
to us.
\paragraph{Kähler-torsion:}
On any almost Hermitian manifold $(M,g,J)$ there exists a unique connection preserving this structure and having totally antisymmetric torsion~\cite{FriedrichIvanov}.
This connection is called the \emph{Kähler-torsion} (KT) \emph{connection} or \emph{Bismut 
connection}~\cite{Bismut:1989}.
KT 6-manifolds are characterized by its torsion, which
is given by
\begin{equation}
 T = J\, \diff \o 
\end{equation}
and which is the real part of a $(2,1)$-form.
From~\cite{FriedrichIvanov} one can see that KT manifolds are complex manifolds, i.e.~they enjoy
\begin{align}
 W_1^\pm = W_2^\pm=0\; .
\end{align}
Note that in general their structure group is U$(3)$ rather than SU$(3)$, as they are a subclass of almost Hermitian structures.
However, they may reduce to an SU$(3)$-structure that is contained in the U$(3)$-structure.
\paragraph{Calabi-Yau-torsion:}
If the KT connection is traceless, its holonomy is SU$(3)$ instead of U$(3)$ 
and, therefore, the structure group is reduced to SU$(3)$.
Conversely, if one is given an SU$(3)$-structure $(g,\omega,\Omega)$ on $M^6$, 
this is always contained in the almost Hermitian structure defined by 
$(g,\omega)$.
The KT connection of the latter then comprises an SU$(3)$ connection for the SU$(3)$-structure if and only if its U$(1)$ part vanishes on the SU$(3)$ subbundle.
This can be written as a further condition on their torsion classes of the 
SU$(3)$-structure under consideration (see, e.g.~\cite{Harland:2010ix}), which 
reads
\begin{align}
 2\,W_4 + W_5=0\; ,
\label{eqn:def_CYT}
\end{align}
without restricting $W_3$.
SU$(3)$-structures that are compatible with the KT connection of their almost Hermitian structure in this sense are called \emph{Calabi-Yau-torsion} (CYT).
Hence, CYT manifolds form a subset of KT manifolds, but with SU$(3)$ structure group.
\paragraph{Nearly Kähler:} An SU$(3)$-structure on a $6$-manifold is \emph{nearly Kähler} if 
\begin{align}
 W_1^+ = W_2^\pm = W_3 = W_4 = W_5 = 0 \;.%
\label{eqn:def_NK}
\end{align}
Note that, in general, one does not need a vanishing $W_1^+$, but this can be 
achieved by suitable phase-transformation in $\Om$.

\paragraph{Half-flat:} An SU$(3)$-structure on a $6$-manifold which satisfies
\begin{align}
 W_1^+ = W_2^+ = W_4 = W_5 =0
\end{align}
is called \emph{half-flat}.

Note that generic nearly Kähler and half-flat $6$-manifolds have a
non-integrable
almost complex structure $J$ and that nearly Kähler manifolds are a subclass of
half-flat manifolds.

%% file: Cones_and_SineCones.tex
\section{Cylinders and sine-cones over 5-manifolds with SU(2)-structure}
\label{sec:Cones_and_SineCones}%
\noindent
Cylinders, metric cones, and sine-cones represent a tool for constructing
$(n{+}1)$-dimensional $G'$-structure manifolds starting from $n$-dimensional
$G$-structure manifolds with $G \subset G'$. At first, we review the well-known
Calabi-Yau cone and the previously presented Kähler-torsion sine-cone~\cite{Bunk:2014kva} for
completeness. Next, we focus on the nearly Kähler sine-cone and the
half-flat cylinder, which will provide the stage for the instanton equations considered in this paper.

First, let us assume we are given a $5$-dimensional manifold $M^5$ with an
SU$(2)$-structure defined by $(\eta,\omega^\a)$ and a Riemannian metric $g_5$.
These tensor fields induce global tensor fields on the Cartesian product $M \times I$, where $I$ is an interval.
Due to the properties~\eqref{eqn:local_SU(2)-forms} of the SU$(2)$-structure on $M^5$, around every point of $M \times I$, there is a local frame such that
\begin{equation}
\label{eqn:SU2_on_MxI}
 \eta = -e^5\;,\quad \omega^\a = \frac{1}{2}\, \eta^\a_{ab}\, e^a \wedge e^b \and
  \diff r = e^6\;,
\end{equation}
if $r$ is the natural coordinate on the interval $I$.
Next, we can apply transformations to these local frames;
for example, perform a transformation like
\begin{subequations}
  \label{eqn:warped_product}%
\begin{equation}%
 e^\mu \mapsto \phi(r)\, e^\mu \and e^6 \mapsto e^6 \;,%
\end{equation}
changing the metric on $M^5 \times I$ to the warped-product metric
\begin{equation}
 g = \diff r^2 + \phi(r)^2\, g_5 \qquad \text{on} \; M^5{\times_\phi}I \;  .%
\end{equation}%
\end{subequations}
Still, the forms $(\phi\, \eta,\, \phi^2\, \omega^\a,\, \diff r)$ will have the
same components as in~\eqref{eqn:SU2_on_MxI} with respect to the altered frames.

Afterwards, one still has the freedom of further transformations. These need to 
map one SU$(2)$-structure to another, which means that the defining forms need 
to have the standard components~\eqref{eqn:local_SU(2)-forms} with respect to 
the new frame. In addition, those transformations can be chosen to preserve the 
warped-product metric. In other words, these admissible transformations are 
given by maps from $M^5
\times I$ to the normalizer subgroup of SU$(2)$ in GL$(6,\mathbb{R})$ (or 
SO$(6)$ if one wants to preserve $g$), i.e.
\begin{equation}
\label{eqn:Trafo_G-Structure}
 L : M^5 \times I \to N_{\mathrm{GL}(6,\mathbb{R})}(\mathrm{SU}(2)) \; .
\end{equation}
The crucial statement is that if we are given a set of forms $(\eta,\omega^\a)$
on $M^5 \times I$ such that around every point in $M^5 \times I$ there is a
local frame with respect to which ~\eqref{eqn:SU2_on_MxI} holds true, the forms
defined by
\begin{subequations}
\label{eqn:SU2_to_SU3}
 \begin{align}
  \omega ={}& \omega^3 - \eta \wedge \diff r\;,\\
  \Omega^+ ={}& - \omega^1 \wedge \diff r + \omega^2 \wedge \eta\;,\\%
  \Omega^- ={}& - \omega^2 \wedge \diff r - \omega^1 \wedge \eta
 \end{align}
\end{subequations}
take the standard components~\eqref{eqn:SU3_standard_components} with respect to
these local frames and, therefore, define an SU$(3)$-structure on $M^5 \times
I$.
Note that $\omega$ and $\Omega$ are globally well-defined, simply because $\eta$ and the $\omega^\a$ are.

This provides us with a general way to construct SU$(3)$-structure manifolds in $6$ dimensions.
Namely we push a given SU$(2)$-structure on $M^5$ forward to $M^5 \times I$ and apply transformations such that we still are given forms with components~\eqref{eqn:SU2_on_MxI}.
Then we know that there exists an extension to an SU$(3)$-structure given
by~\eqref{eqn:SU2_to_SU3}.
In the following subsections we apply this procedure in several cases.
\subsection{Calabi-Yau metric cones}
One result that makes Sasaki-Einstein manifolds interesting for string theorists
as well as mathematicians is that their metric cones are Calabi-Yau. Here we
demonstrate this explicitly for the $5$-dimensional case. Consider a
Sasaki-Einstein $5$-manifold $M^5$ with local coframes $e^\mu$, 
where $\mu = (a,5)$ and $a=1,2,3,4$. The metric on its metric cone reads
\begin{equation}
 g = r^2 \left( \delta_{ab}\, e^a \otimes e^b + e^5 \otimes e^5 \right) +  \diff r
\otimes \diff r  = r^2 \left( \delta_{ab}\, e^a \otimes e^b + e^5 \otimes e^5 + e^6
\otimes e^6\right) \label{eqn:metric_Calabi-Yau} 
\end{equation}
with 
\begin{align}
 e^6 = \diff\tau = \frac{\diff r }{r} \; . 
\end{align}
The last equality in \eqref{eqn:metric_Calabi-Yau} displays the conformal
equivalence to the cylinder over $M^5$ with the metric
\begin{align}
 g_\text{cyl}=\de_{ab}\, e^a \otimes e^b + e^5 \otimes e^5 + e^6 \otimes e^6 \;
.%
\label{eqn:metric_cylinder}
\end{align}
We can introduce an almost complex structure $J$ on the metric cone via
\begin{align}
 J \hat{\Theta}^\a = \im \hat{\Theta}^\a \for \a=1,2,3 \with \hat{\Theta}^\a =
\hat{e}^{2\a -1} + \im \hat{e}^{2\a}\;,
\end{align}
and we set $\hat{e}^{\hat{\mu}} = r e^{\hat{\mu}}$ for $\hat{\mu}= 1, \ldots,
6$. The SU$(3)$-structure forms $(\hat{\o},\hat{\Om})$ have the local expressions
\begin{subequations}
\begin{align}
 \hat{\o} &= \hat{e}^{1} \wedge \hat{e}^{2} + \hat{e}^{3} \wedge \hat{e}^{4} +
\hat{e}^{5} \wedge \hat{e}^{6} = r^2 (\o^3  + e^5 \wedge e^6) \; ,\\
\hat{\Om} &= \hat{\Theta}^1 \wedge \hat{\Theta}^2 \wedge \hat{\Theta}^3 \; ,
\end{align}%
\end{subequations}%
for which a direct computation yields
\begin{align}
 \diff \hat{\o} =0 \und \diff \hat{\Om} =0 \; .
\end{align}
Therefore, the metric cone introduced in~\eqref{eqn:metric_Calabi-Yau} is indeed Calabi-Yau as all SU$(3)$-torsion classes vanish.
\subsection{Kähler-torsion sine-cones}
\label{subsec:KT-SineCone}
Consider a Sasaki-Einstein 5-manifold $M^5$ and the product manifold
$M^6 = M^5 \times \left(0,\La \pi\right)$ with the metric 
\begin{subequations}
\begin{align}
g ={}& \La^2 \sin^2\!\varphi\left(\de_{ab}\, e^a \otimes e^b + e^5 \otimes e^5
\right)  +  \diff r \otimes \diff r \\%
={}& \La^2 \sin^2\!\varphi \left( \de_{ab}\, e^a \otimes e^b + e^5 \otimes e^5 +
 e^6 \otimes e^6 \right)\;,%
\label{eqn:metric_KTSineCone}
\end{align}
\end{subequations}
where 
\begin{align}
 \varphi = \frac{r}{\La} \and e^6 = \diff \tau = \frac{\diff \varphi}{\sin\varphi} \; ,
\end{align}
and $\La \in \mathbb{R}^+$ is a scaling parameter. Equation 
\eqref{eqn:metric_KTSineCone} shows that
the metric on the sine-cone is conformally equivalent to the
metric~\eqref{eqn:metric_cylinder} on the cylinder over $M^5$.

The explicit solution of $\tau = \tau (\varphi)$ is computed to
\begin{align}
\tau = \ln \left|\tan\sfrac{\varphi}{2}\right| +
\text{constant} \; ,
\end{align}
and the integration constant can be chosen such that the sine-cone becomes the 
metric cone in the limit $\La \to \infty$. Hence, the computation yields
\begin{align}
 \tau (\varphi) =  \ln \left( 2 \La \tan\sfrac{\varphi}{2} \right)
 = \ln \left( 2 \La \sqrt{ \sfrac{1-\cos\varphi}{1+\cos\varphi} } \right) \; . %
\label{eqn:relation_tau_phi}
\end{align}

Next, we introduce an almost complex structure $J$ and the associated
fundamental $(1,1)$-form $\tilde{\o}$ on the sine-cone as follows ($\a = 
1,2,3$):
\begin{subequations}
\label{eqn:compl-structure-KT}
\begin{alignat}{2}
 J \tilde{\Th}^\a &= \im  \tilde{\Th}^\a & \with \tilde{\Th}^\a &=
\La \sin \varphi \left( e^{2 \a - 1} + \im e^{2 \a } \right) \; ,\\
J \tilde{\Th}^{\bar{\a}} &= -\im  \tilde{\Th}^{\bar{\a}} & \with
\tilde{\Th}^{\bar{\a}} &=  \overline{\tilde{\Th}^\a}  \; ,\\
\tilde{\o} &= \La^2 \sin^2\!\varphi \left( \o^3 + e^5 \wedge e^6 \right) \; , & 
&
\end{alignat}
\end{subequations}
where $\o^3$ is defined in \eqref{eqn:local_SU(2)-forms}.
As shown in~\cite{Bunk:2014kva}, the above structure comprises a Kähler-torsion structure on the sine-cone.
That is, there exists the uniquely defined Bismut $\nabla^B$ connection, which preserves $g$ and $J$, and has torsion given by
\begin{equation}
 T^B = J\, \diff \tilde{\o}\;.
\end{equation}

\paragraph{Remarks:}
One can also introduce a globally well-defined complex $(3,0)$-form $\tilde{\Om}$ defined as
\begin{align}
 \tilde{\Om} = \tilde{\Th}^1 \wedge \tilde{\Th}^2 \wedge \tilde{\Th}^3
  = \Lambda^3 \sin(\varphi)^3 \big( \omega^2 - \im \omega^1 \big) \wedge \eta
  - \Lambda^2 \sin(\varphi)^2 \big( \omega^1 + \im \omega^2 \big) \wedge \diff r \; .
 \label{eqn:Omega-KT}
\end{align}
Applying the exterior differential yields
\begin{subequations}
\begin{align}
 \diff \tilde{\o}&= 2 \, \frac{ \cos\varphi -1}{\La \sin\varphi}\,\tilde{\o} 
\wedge \tilde{e}^6
 = -\frac{2}{\La} \tan\sfrac{\varphi}{2}\,\tilde{\o} \wedge \tilde{e}^6 \; ,\\
 \diff \tilde{\Om} &= 3 \, \frac{1-\cos\varphi}{\La \sin\varphi}\, \tilde{\Om} 
\wedge \tilde{e}^6
 = \frac{3}{\Lambda}\, \tan\sfrac{\varphi}{2}\, \tilde{\Om} \wedge 
\tilde{e}^6 \; ,
\end{align}%
 \label{eqn:KT-exteriorDiffs}%
\end{subequations}%
thus rendering the sine-cone over $M^5$ an SU$(3)$-structure manifold as defined in Section~\ref{sec:SU3_structures}.
From \eqref{eqn:KT-exteriorDiffs} we immediately see that $J$ is
integrable and 
\begin{align}
 2 W_4 + W_5 = -\frac{1}{\La} \tan\sfrac{\varphi}{2}\,\tilde{e}^6
\neq 0 \; \text{ for } \; \La < \infty \; ,
\end{align}
whence the Bismut connection does not preserve the SU$(3)$-structure unless $\La=\infty$.
Nevertheless, the condition $3\, W_4 + 2\, W_5 = 0$ is satisfied, which is in 
agreement with the conformal equivalence between the sine-cone over a 
Sasaki-Eintein $5$-manifold and the Calabi-Yau metric cone over 
$M^5$~\cite{Chiossi:2002tw,LopesCardoso:2002hd}.
That is, the conformal equivalence of the Calabi-Yau cone and the Kähler torsion 
sine-cone also maps their two SU$(3)$-structures onto one another.
We also note that $2 W_4 + W_5 \to 0$ as $\La \to \infty$, and the KT sine-cone becomes
the Calabi-Yau metric cone.
Recall from section~\ref{sec:SU3_structures} that Kähler-torsion structures are U$(3)$-structures, whence one has to distinguish between this and the additional SU$(3)$-structure.
\subsection{Nearly Kähler sine-cones}
\label{subsec:NK-SineCone}
In~\cite{Fernandez:2006ux} a nearly Kähler structure on the sine-cone over a Sasaki-Einstein 5-manifold has been obtained by means of flow equations.
Here, in contrast, we show that this structure can be constructed by means
of a combined rotation and rescaling of the coframes of the cylinder over the
Sasaki-Einstein $5$-manifold.
We will carry this construction out in the following three steps:
\begin{enumerate}
  \item An SU$(3)$-structure on the cylinder over a Sasaki-Einstein 5-manifold
$M^5$ can be introduced via a metric~\eqref{eqn:metric_cylinder}, an almost
complex structure $J$ or the equivalent $(1,1)$-form $\o$, and a $(3,0)$-form
$\Om$. These objects are 
\begin{subequations}%
\label{eqn:SU(3)_cylinder}%
\begin{align}%
 \o&= \o^3 + e^5 \wedge e^6 = e^1 \wedge e^2 + e^3 \wedge e^4 + e^5 \wedge e^6
\; , \\%
J \Th^\a &= \im \Th^\a \quad \text{ for } \quad \Th^{\a} = e^{2\a-1} + \im e^{2\a}
\quad\text{with}\quad \a=1,2,3 \; , \\%
\Om &= \Th^1 \wedge \Th^2 \wedge \Th^3 = - \o^2 \wedge e^5 - \o^1 \wedge e^6 +
\im \left( \o^1 \wedge e^5 - \o^2 \wedge e^6 \right) \; .%
\end{align}%
\end{subequations}%
  \item Next, we consider an SO$(5)$-rotation of the SU$(2)$-structure $(\h, \o^\a)$ on $M^5$.
Let $\h^2$ be the matrix of the 't~Hooft symbols $\h^2_{ab}$ and perform a
rotation of the basis $1$-forms $e^1, \ldots , e^4$,
\begin{equation}
 E= \begin{pmatrix} e^1 \\ e^2 \\ e^3 \\ e^4  \end{pmatrix} \quad\mapsto\quad E_\varphi =
\exp{\left( \sfrac{\varphi}{2}\, \h^2 \right)}\, E = %
\begin{pmatrix} \cos \frac{\varphi}{2} & 0 & -\sin\frac{\varphi}{2} & 0 \\ %
0 &  \cos\frac{\varphi}{2}& 0 & \sin\frac{\varphi}{2}\\%
\sin\frac{\varphi}{2} & 0 & \cos\frac{\varphi}{2} & 0 \\ %
0 & - \sin\frac{\varphi}{2} & 0 & \cos\frac{\varphi}{2} \end{pmatrix} 
\begin{pmatrix} e^1 \\ e^2 \\ e^3 \\ e^4  \end{pmatrix}\;. %
\label{eqn:rotation_SU(2)-structure}
\end{equation}
In the rotated frame $(e_\varphi^a,e^5)$ we define the SU$(3)$-structure
forms to have the same components as in the unrotated
frame~\eqref{eqn:SU(3)_cylinder}, i.e.
\begin{subequations}
\begin{align}
 \o_\varphi&= \o_\varphi^3 + e^5 \wedge e^6  \; ,\\%
\Om_\varphi &=  - \o_\varphi^2 \wedge e^5 - \o_\varphi^1 \wedge e^6 +
\im \left( \o_\varphi^1 \wedge e^5 - \o_\varphi^2 \wedge e^6 \right) \; ,%
\end{align}%
\label{eqn:SU(3)_rotated-cylinder}%
\end{subequations}%
where $\o_\varphi^\alpha = \frac{1}{2} \eta^\alpha_{\mu \nu} e_{\varphi}^{\mu \nu}$.
Note that this is still an SU$(3)$-structure on the cylinder, because the
defining forms still have the standard components~\eqref{eqn:SU(3)_cylinder}
with respect to the coframes $e_\varphi^\mu$.
\item Last, the pullback to the sine-cone $C_s (M^5)$ along the map establishing the conformal
equivalence to the cylinder yields
\begin{subequations} 
\begin{alignat}{2} 
 e_s^a &= \La \, e_\varphi^a \,\sin\varphi\;  , & \quad  e_s^5 &= \La \, e^5 \,\sin\varphi\; ,
 \qquad e_s^6 = \La \, e^6 \,\sin\varphi = \La\, \diff \varphi = \diff r \; ,\\%
 \o_s^\a &= \La^2 \, \o_\varphi^\a \,\sin^2\!\varphi \; , & \; \o_s &= \o_s^3 + 
\La^2\, e^5\wedge e^6\,\sin^2\!\varphi \; ,\\%
 \Om_s &= \La^3 \, \Om_\varphi \,\sin^3\!\varphi& & %
\end{alignat}%
\label{eqn:SU(3)_nearlKähler}%
\end{subequations}%
as an SU$(3)$-structure on the sine-cone.
By a direct calculation we obtain
\begin{subequations}
 \begin{align}
  \diff \o_s &= -\frac{3}{\La}\, \Om^+_s  \; ,\\
 \diff \Om^+_s &= 0 \; , \qquad \diff \Om^-_s = \frac{2}{\La}\, \o_s \wedge \o_s \; ,
 \end{align}
\end{subequations}
which confirms that \eqref{eqn:SU(3)_nearlKähler} induces a nearly Kähler structure on the sine-cone.
\end{enumerate}

\paragraph{Remarks:} In the limit $\Lambda \to \infty$, in which the sine-cone
becomes
the metric cone, this nearly Kähler structure on the sine-cone is
smoothly deformed to the Calabi-Yau structure on the metric cone since
\begin{align}
 \lim _{\La \to \infty} \diff \o_s = 0 \und \lim _{\La \to \infty} \diff \Om_s = 0 \; .
\end{align}

Generically, the sine-cone, as a conifold, has two
singularities at $\varphi=0$ and $\varphi=\pi$. As we see
from~\eqref{eqn:SU(3)_nearlKähler}, the SU$(3)$-structure cannot be extended 
to the tips, because all defining forms vanish at these points. Hence, the
sine-cone is a nearly Kähler manifold only for $\varphi \in (0,\pi)$, and one
cannot add the singular points.
\subsection{Half-flat cylinders}
\label{subsec:HF-Cylinder}
Consider a $5$-dimensional manifold $M^5$ endowed with a Sasaki-Einstein
SU$(2)$-structure defined by $(\eta,\omega^1,\omega^2,\omega^3)$ as in
Section~\ref{sec:SU(2)-structures}.
For an arbitrary coframe $e^\mu$ belonging to the SU$(2)$-structure, consider 
the transformation
\begin{subequations}
\label{eqn:trafo_half-flat}
 \begin{alignat}{2}
  e_z^1 ={}& e^4 \cos\zeta + e^3 \sin\zeta \;,&  \quad  e_z^2 ={}& -e^1  \;,\\%
  e_z^3 ={}& e^2\;, & \quad  e_z^4 ={}& e^3 \cos\zeta - e^4 \sin\zeta\;,\\%
  e_z^5 ={}& \varrho\, e^5. & &%
 \end{alignat}
\end{subequations}
Here $\zeta \in [0,2\pi]$ and $\rho \in \mathbb{R}^+$ are two constant parameters.
For $\varrho = 1$ this can be seen to be an SO(5)-transformation of the coframe, such that the metric on $M^5$ is unchanged.
Nevertheless, we obtain a two-parameter family of SU$(2)$-structures on $M^5$ by
defining
\begin{equation}
 \eta_z = \varrho\, \eta, \quad \omega_z^\alpha = \sfrac{1}{2}\, \eta^\alpha_{\mu \nu}\, e_z^\mu \wedge e_z^\nu, \quad g_z = \delta_{\mu \nu}\, e_z^\mu \otimes e_z^\nu\; .
\end{equation}
These are globally well-defined as can be seen from
\begin{subequations}
 \begin{align}
  \omega_z^1 ={}& - \omega ^3  \;,\\%
  \omega_z^2 ={}& \omega^1\,\sin\zeta + \omega^2\,\cos\zeta\;,\\%
  \omega_z^3 ={}& \omega^1\,\cos\zeta - \omega^2\,\sin\zeta\;,%
 \end{align}%
\end{subequations}%
and, thus, yield a two-parameter family of SU$(2)$-structures on $M^5$.
Note that these structures are neither hypo nor nearly hypo any more.

With these SU$(2)$-structures on $M^5$ at hand we define a two-parameter family
of SU$(3)$-structures on the metric cylinder $(M^5 \times \mathbb{R},\, \bar{g}_z
= g_z + \diff r \otimes \diff r)$ by
\begin{subequations}
 \begin{align}
  \omega_z = {}& \omega_z^3 - \eta_z\wedge \diff r
   = \omega^1\,\cos\zeta - \omega^2\,\sin\zeta - \varrho\, \eta
\wedge \diff r \; ,\\%
  \Omega_z^+ ={}& - \omega_z^1 \wedge \diff r + \omega_z^2 \wedge 
\eta_z    = \varrho\, \big( \omega^1\,\sin\zeta + \omega^2\,\cos\zeta
\big) \wedge \eta +  \omega^3 \wedge \diff r  \;,\\%
  \Omega_z^- ={}& - \omega_z^2 \wedge \diff r - \omega_z^1 \wedge
\eta_z  =  -\big( \omega^1\,\sin\zeta + \omega^2\,\cos\zeta\big)
\wedge \diff r + \varrho\, \omega^3 \wedge \eta  \;,
 \end{align}
\end{subequations}
which yields a two-parameter family of half-flat SU$(3)$-structures.
The non-vanishing torsion classes can be computed to read
\begin{equation}
\label{eqn:torsion_hfcylinder}
\begin{aligned}
 W_1^- &= \frac{3 + 2 \varrho^2}{3 \varrho} \; ,
 \quad W_2^- = \frac{4 \varrho^2 - 3 }{3 \varrho}\, \left( \omega_z^3 + 2\,
\eta_z \wedge \diff r \right)\and \\%
 \quad W_3 &= \frac{2 \varrho^2 - 3}{2 \varrho} \left(
\omega_z^1 \wedge \diff r  + \omega_z^2 \wedge \eta_z \right)\;.
\end{aligned}
\end{equation}
Furthermore, the conditions $\o_z \lrcorner W_2^-=0$ and $\o_z
\lrcorner W_3=0$ are satisfied for any values of the parameters $\zeta$ and
$\varrho$.
\subsection{Summary of cone constructions}
The different cone constructions linking Sasaki-Einstein to U$(3)$ or SU$(3)$ 
$6$-manifolds, which have been presented 
in~\cite{Bunk:2014kva} and this paper, are summarized in the following 
table: 
\noindent
\begin{table}[htpb!]
\begin{tabularx}{\textwidth}{|p{3.1cm}|p{3.8cm}|p{3.2cm}|X|}
 \hhline{*{4}{|-}|}
 \textbf{structure on $M^5$} & \textbf{cone construction} & \textbf{structure on $M^6$} & \textbf{non-zero torsion classes}\\ \hhline{*{4}{|=}|}
 Sasaki-Einstein & cone & Calabi-Yau & $--$ \\ \hhline{|~|---|}
  & sine-cone & Kähler-torsion & \\ \hhline{|~|---|}
  & sine-cone with rotation & nearly Kähler & $W_1^-$
\\ \hhline{|~|---|}
  & cylinder with rotation & half-flat & $W_1^-,W_2^-,W_3$ \\ \hhline{*{4}{|-}|}
  \end{tabularx}
\caption{\textsl{Summary of cone constructions linking Sasaki-Einstein to 
U$(3)$ or SU$(3)$-structures in $d=6$ and the non-zero torsion classes for the respective SU$(3)$-structures.}}
\end{table}

%% file: instantons_6-manifolds.tex
\section{Instantons on conical 6-manifolds}
\label{sec:Instantons}
\subsection{Definition and reduction of instanton equations on conical 6-manifolds}
\label{sec:InsttoMatrix}
Having constructed several $6$-dimensional SU$(3)$ manifolds in the last section, we now turn our attention to instanton equations on such spaces.
Thus, let $M^6$ be a 6-manifold with a connection $\mathcal{A}$ on the tangent
bundle. The
curvature $2$-form $\mathcal{F}$ associated to $\mathcal{A}$ is given by 
\begin{align}
 \mathcal{F} = \diff \mathcal{A} + \mathcal{A} \wedge \mathcal{A} \eqqcolon D_\mathcal{A} \mathcal{A} \; ,
\end{align}
where $D_\mathcal{A}$ is the covariant differential associated to $\mathcal{A}$, and the Bianci identity $D_\mathcal{A} \mathcal{F} = 0$ holds true.
As before, we can perform the type-decomposition of a form with respect to any almost complex structure $J$, yielding
\begin{align}
 \mathcal{F} = \mathcal{F}^{2,0} + \mathcal{F}^{1,1} + \mathcal{F}^{0,2} \; .
\end{align}
For a given SU$(3)$-structure $(\o, \Om)$ on a $6$-manifold and a curvature
 $2$-form $\mathcal{F}$, the instanton equation can be defined in two steps:
first, the \emph{pseudo-holomorphicity condition} reads
\begin{subequations}%
\begin{align}%
 \Om \wedge \mathcal{F} =0 \quad \Leftrightarrow \quad \mathcal{F}^{0,2} = 0 \; ,%
\label{eqn:instanton_1}%
\end{align}%
and, second, applying the covariant differential to \eqref{eqn:instanton_1}, and
using the Bianchi identity as well as \eqref{eqn:instanton_1}
yields
\begin{align}%
 \diff \Om \wedge \mathcal{F} = \left[ \left( W_1^+ + \im W_1^- \right) \o
\wedge \o + \left( W_2^+ + \im W_2^- \right) \wedge \o \right] \wedge
\mathcal{F} = 0 \; .%
\label{eqn:instanton_2}%
\end{align}%
\label{eqn:def_instanton}%
\end{subequations}%
The last equation, although a mere consequence of~\eqref{eqn:instanton_1},  
depends strongly on the type of SU$(3)$-manifold under consideration.
For example, on nearly Kähler manifolds one has
\begin{align}
 \diff \Om \propto \o \wedge \o \qquad  \xRightarrow{\eqref{eqn:instanton_2} \;}
\qquad \o \wedge \o \wedge \mathcal{F} = 0 \quad\Leftrightarrow\quad \o \lrcorner \mathcal{F} =0 \; ,
\end{align}
whereas on half-flat SU$(3)$-manifolds this is not true as $\diff \Om \neq \kappa\, \o \wedge \o$.
For Calabi-Yau spaces, on the other hand, \eqref{eqn:instanton_2} is trivial as $\diff \Om=0$, and the condition $\o \lrcorner \mathcal{F} = 0$ is added as an additional stability condition for the holomorphic instanton bundle \cite{Donaldson01011985,donaldson1987,Uhlenbeck:CPA1986}.

Following \cite{Ivanova:2012vz}, one considers a complex vector bundle $\mathcal{V} \to M^6$ of rank $k$ on which we are given an instanton $\Gamma$ with curvature $R_\Gamma$.
Here this vector bundle will be the tangent bundle of $6$-manifolds arising as certain conical extensions of SU$(2)$ $5$-manifolds $M^5$, just as we considered in the previous section.
We then generalize this instanton $\Gamma$ by extending it to a connection $\mathcal{A}$ with curvature $\mathcal{F}$ by the ansatz
\begin{align}
 \mathcal{A} = \Gamma + X_\mu e^\mu  \und \mathcal{F} = \diff \mathcal{A} + \mathcal{A} \wedge \mathcal{A} \; , %
\label{eqn:ansatz_generic}
\end{align}
where $\mu= 1,\ldots,5$ and
\begin{align}
 \Gamma = \Gamma^i \hat{I}_i \; , \quad i = 6,7,8\;.
\end{align}
Here $\hat{I}_i$ is a representation of the
SU$(2)$-generators $I_i$ on the fibres $\mathbb{R}^6$ of the bundle, and
$\Gamma^i$ are the components of an $\mathfrak{su}(2)$-connection
on the tangent bundle of $M^6$. Furthermore,
$X_\mu$ are matrices from $\mathrm{End}(\mathbb{R}^6)$.

The computation of $\mathcal{F}$ with the ansatz for $\mathcal{A}$ yields
\begin{equation}
\begin{aligned}
 \mathcal{F} = R_\Gamma &+ \diff X_\mu \wedge e^\mu + T_{6 \nu}^{\mu} X_\mu e^6
\wedge e^\nu + \sfrac{1}{2} \left( [X_\mu, X_\nu] + T_{\mu \nu}^\s X_\s \right)
e^\mu \wedge e^\nu \\
&+ \Gamma^i \left( [\hat{I}_i,X_\mu] - f_{i \mu}^\nu  X_\nu
\right) \wedge e^\mu \; . 
\end{aligned}
\label{eqn:generic_curvature}%
\end{equation}
Herein, $T$ denotes the torsion of the connection $\Gamma$.

In order to simplify this further, we investigate the matrices $X_\mu$ and their 
transformation behavior under a change of $e$.
By construction, $X_\mu e^\mu$ is the local representation of an $Ad$-equivariant $1$-form $X$ on the gauge principal bundle, which here coincides with the SU$(3)$-subbundle $\mathcal{P}$ of the frame bundle of $M^6$ that constitutes the SU$(3)$-structure.
Note that, in the aforementioned cases, $\mathcal{P}$ contains a principal 
SU$(2)$-subbundle $\mathcal{Q}$; the latter is the principal bundle for the 
connection $\Gamma$. Now let $e$ and $e'$ be two local sections of $\mathcal{Q} 
\subset \mathcal{P}$ over some $U \subset M^6$ related by an 
SU$(2)$-transformation $L: U \to SU(2)$. The components $X'_{\mu}$ and $X_\mu$ 
of $X$ with respect to $e'$ and $e$ are related via
\begin{equation}
\label{eqn:local trafo X}
 X'_{\mu} = Ad(L^{-1}) \circ X_\nu\, \rho(L)^\nu_{\mu}\; .
\end{equation}
Here $\rho$ is the representation of SU$(2)$ on $\mathbb{R}^5$ which is the
typical fiber of $TM^5$. It coincides with the representation $Ad_{SU(3)}: SU(2)
\to \mathrm{End}(\mathfrak{m})$, where $\mathfrak{su}(3) = \mathfrak{su}(2)
\oplus \mathfrak{m}$ and one has the identification $\mathfrak{m} \simeq T_x
M^5$.

Since we will search for $\mathfrak{su}(3)$-valued connections $\mathcal{A}$, we
consider the $\mathfrak{su}(3)$-generator algebra
\begin{subequations}
\begin{alignat}{2}
 [ \hat{I}_i , \hat{I}_j ] &= f_{i j}^{\ \ k} \hat{I}_k \; , & \qquad
i,j,k &= 6,7,8 \\
 [ \hat{I}_i , \hat{I}_\mu ] &= \ddu{f}{i}{\mu}{\nu} \hat{I}_\nu \; , &
\mu,\nu,\sigma &=
1,2,3,4,5 \\
 [ \hat{I}_\mu , \hat{I}_\nu ] &= \ddu{f}{\mu}{\nu}{i} \hat{I}_i +
\ddu{f}{\mu}{\nu}{\s} \hat{I}_\s ; . & &
\end{alignat}
\end{subequations}
The generators with indices $i,j,k$ belong
to the $\mathfrak{su}(2)$ subalgebra, and the indices $\mu,\nu,\s$ correspond to
its orthogonal complement $\mathfrak{m}$ in the
SU$(2)$-invariant splitting
\begin{align}
 \mathfrak{su}(3) = \mathfrak{su}(2) \oplus \mathfrak{m} \; . %
\label{eqn:splitting_LieAlgebra}
\end{align}

Generically, only $X$ is well-defined globally, rather than the component maps
$X_\mu$.
The latter strongly depend on the choice of the local frame $e$ and, therefore,
we have no control over their behavior in general.
That would be different, if the components $X_\mu$ were independent of
the trivialization of the involved bundles, that is, if the $X_\mu$ were
invariant under the aforementioned transformations~\eqref{eqn:local trafo X}
that change the local frames.
Furthermore, since SU$(2)$ is connected, this is equivalent to the infinitesimal 
version of the invariance, i.e.
\begin{equation}
 [\hat{I}_i,X_\mu] = \rho_*(I_i)^\nu_\mu\, X_\nu = \ddu{f}{i}{\mu}{\nu}\, X_\nu \;. %
 \label{eqn:equivariance}%
\end{equation}
Note that this simplification implies that 
the $X_\mu$ are independent of the choice of frame adapted to the 
SU$(2)$-structure $\mathcal{Q}$; hence, we can choose them to vary with 
the cone direction only. Condition~\eqref{eqn:equivariance} appeared, for 
example, in \cite{Kobayashi:Vol2,Kapetanakis:1992} on coset spaces,
where equivariant
connections have been constructed.
We will in the following refer to
\eqref{eqn:equivariance} as the \emph{equivariance condition}, despite its 
different origin in this context. 

Inserting this simplification and the accompanying consistency condition
\eqref{eqn:equivariance} into \eqref{eqn:generic_curvature}, we are left with
\begin{equation}
 \mathcal{F} = R_\Gamma + \big( \dot{X}_\mu\, + T^\nu_{6\mu}\, X_\nu \big)\, e^6
\wedge e^\mu + \sfrac{1}{2} \left( [X_\mu ,  X_\nu] + T_{\mu \nu}^\s X_\s
\right) e^\mu \wedge e^\nu\; .
\end{equation}
Here the dot denotes the derivation in the cone direction.
In any case, the instanton condition is the requirement that the $2$-form part
of
$\mathcal{F}$ takes values in a certain subbundle of $\Lambda^2 T^*M^6$,
which we call the \emph{instanton bundle}.
Anticipating that $2$-forms of the general form $e^6 \wedge e^\sigma +
\frac{1}{2} N^\sigma_{\mu \nu} e^\mu \wedge e^\nu$, with $N$ to be determined
from the geometry under consideration, are local sections of this instanton bundle, we add a zero to the above expression and obtain
\begin{equation}
\label{eqn:F(canonical)withN}
 \begin{aligned}
  \mathcal{F} ={}& R_\Gamma + \big( \dot{X}_\mu\, + T^\nu_{6\mu}\, X_\nu \big)\,
\big( e^6 \wedge e^\mu + \sfrac{1}{2} N^\mu_{\sigma \rho}\, e^\sigma \wedge
e^\rho \big)\\
  & + \sfrac{1}{2} \left( [X_\mu , X_\nu] + T_{\mu \nu}^\s\, X_\s - N^\sigma_{\mu \nu}\, \big( \dot{X}_\sigma\, + T^\rho_{6\sigma}\, X_\rho \big)\, \right) e^\mu \wedge e^\nu\; .%
 \end{aligned}
\end{equation}
As argued above, $R_\Gamma$ and the second term already are instantons.
Thus, we are left to require that the last term satisfies the instanton
equation; this leads us to
\begin{equation}
 [X_\mu , X_\nu] + T_{\mu \nu}^\s X_\s  = N^\sigma_{\mu \nu}\, \big( \dot{X}_\sigma\, + T^\rho_{6\sigma}\, X_\rho \big) + \mathcal{N}_{\mu \nu} \; ,
\end{equation}
where $\mathcal{N}$ has to be an instanton on $M^6$ that compensates for the 
$\mathfrak{su}(2)$-component of the left-hand-side commutator.
Hence, $\mathcal{N}$ can only be a
linear combination of the three instantons~\cite{Harland:2011zs}
$\ddu{f}{\mu}{\nu}{i} e^\mu \wedge e^\nu$ for $i = 6,7,8$, which depends on the cone coordinate.
That is,
\begin{equation}
 [X_\mu , X_\nu] + T_{\mu \nu}^\s X_\s  = N^\sigma_{\mu \nu}\, \big( \dot{X}_\sigma\, + T^\rho_{6\sigma}\, X_\rho \big) + \ddu{f}{\mu}{\nu}{i}\,\mathcal{N}_i \; . %
\label{eqn:parametrise_X}%
\end{equation}
In summary, we are searching for $\mathfrak{m}$-valued matrices $X_\mu$ that solve equations \eqref{eqn:equivariance} and \eqref{eqn:parametrise_X}, as these will give rise to new instantons on the considered manifolds.
\subsection{Remarks on the instanton equation}
\label{subsec:remarks_Instanton}
Before proceeding with the particular cases of the nearly Kähler sine-cone and
the half-flat cylinder, one needs to clarify an important point regarding the
transformations of coframes mentioned in
Section~\ref{sec:Cones_and_SineCones}.

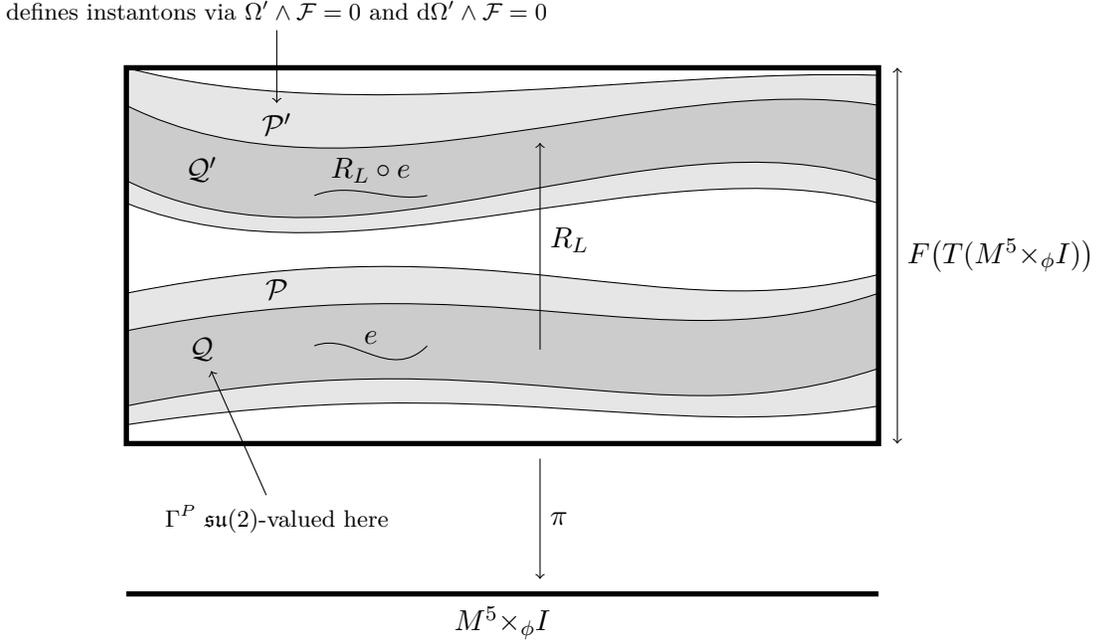
\begin{figure}[h!]
 \input{diagramm}%
 \caption{\textsl{A schematic depiction of the 
different principal bundles
involved in the definition of the instanton condition: $\mathcal{Q}$ and
$\mathcal{P}$ are the SU$(2)$- and SU$(3)$-bundles, respectively, which
originate from the Sasaki-Einstein structure on $M^5$. The transformation $L$ 
defines the principal bundles $\mathcal{Q}'$ and $\mathcal{P}'$, which again 
are SU$(2)$ and SU$(3)$-bundles, respectively. All bundles under
consideration are understood as principal subbundles of the frame 
bundle $F(T(M^5{\times_\phi}I))$.}} 
 \label{fig:G-bundles}
\end{figure}

The SU$(2)$-structure on the Sasaki-Einstein $5$-manifold is understood as an
SU$(2)$-principal bundle $\mathcal{Q}$, a subbundle of the frame bundle
$F(T M^5)$. The warped product $M^5{\times_{\phi}}I$
(c.f.~\eqref{eqn:warped_product}) is equipped
with an SU$(3)$-structure via~\eqref{eqn:SU2_to_SU3} and the corresponding
principal bundle is denoted with $\mathcal{P} \subset 
F(T(M^5{\times_{\phi}}I))$ (c.f. Fig.~\ref{fig:G-bundles}). However,
$\mathcal{P}$ is not the
SU$(3)$-structure one is
interested in, i.e. in our cases it is neither nearly Kähler nor half-flat. The
constructions of Subsections~\ref{subsec:NK-SineCone} and
\ref{subsec:HF-Cylinder} rely on transformations of the coframes on
$M^5$: they generate a different SU$(2)$-structure
$\mathcal{Q}'$ that can be extended to the desired SU$(3)$-structure
$\mathcal{P}'$ on the warped product. 
An important observation is the following: for a $G$-structure $\mathcal{Q}$ 
the bundle $\mathcal{Q}'$ defined via $\mathcal{Q}' = R_L 
\mathcal{Q} $ is a $G$-structure if and only if $L$ is a map from the base to 
the normalizer $N_{\mathrm{GL}(6,\mathbb{R})}(G)$, 
c.f.~\eqref{eqn:Trafo_G-Structure}.

The crux of the instanton equation is the following: the defining forms
$(\o',\Om')$ stem from $\mathcal{P}'$, whereas the canonical connection
$\Gamma^P$ belongs to $\mathcal{Q}$ and is trivially lifted to an instanton on
$\mathcal{P}$. Let us denote by $e\in \Gamma(U,\mathcal{Q})$ an adapted frame 
for $\mathcal{Q}$. Then by construction $e'\eqqcolon (R_L \circ e) \in 
\Gamma(U,\mathcal{Q}')$ is an adapted frame for $\mathcal{Q}'$. By standard 
results, the connection $1$-forms of $\mathcal{A}$ transform under a 
change of section as
\begin{equation}
 {e'}^{\ast} \mathcal{A} = \mathrm{Ad}(L^{-1}) \circ e^\ast \mathcal{A} +
L^{-1} \diff L \; . %
\label{eqn:trafo_connection}
\end{equation}
The employed extension $\mathcal{A}= \Gamma^P + X$ relies on the
splitting~\eqref{eqn:splitting_LieAlgebra} such that $X$ corresponds to
$\mathfrak{m}$-valued $1$-forms. However, this only holds in the frame $e$, due
to the following: Starting with $\Gamma^P$ on $\mathcal{Q}$, one has a purely
$\mathfrak{su}(2)$-valued connection. Applying any transformation $L$ to
$\mathcal{Q}$, $\Gamma^P$ is generically not an $\mathfrak{su}(2)$-valued 
connection on $\mathcal{Q}'$. This is due to the fact that $L^{-1} 
\diff L$, in general, takes values in the Lie-algebra of 
$N_{\mathrm{GL}(6,\mathbb{R})}(\mathrm{SU}(2))$ instead of 
$\mathfrak{su}(2)$. Therefore, one cannot simply take $ 
{e'}^{\ast} \Gamma^P $ as a starting point for some ansatz like $ {e'}^{\ast} 
\mathcal{A} = {e'}^{\ast} \Gamma^P + {X'}_\mu {e'}^\mu$.

For the cases under consideration, $L$ depends (at
most) on the cone direction $r$. Hence, one has that $
\mathrm{Ad}(L^{-1}) \circ e^\ast \mathcal{A} $ is $\mathfrak{su}(2)$-valued and
$L^{-1} \diff L \propto \diff r$, but generically not $\mathfrak{su}(2)$-valued.
The immediate consequences are the following:
\begin{itemize}
\item For instance, on the nearly Kähler sine-cone one has to perform all
calculations in the frame $e$, because for the 
derivation of Subsection~\ref{sec:InsttoMatrix} we employed a section of the 
bundle on which $\Gamma^P$ is an $\mathfrak{su}(2)$-valued connection. We will, 
however,
compute ${e'}^\ast \Gamma^P$ explicitly in Subsection~\ref{subsec:CanonConSU(3)}
and demonstrate that it yields an $\mathfrak{su}(3)$-valued instanton on the
sine-cone.
\item  In contrast, the transformation
for the half-flat cylinder~\eqref{eqn:trafo_half-flat} is, although a
$2$-parameter family, base-point independent. Therefore, one is allowed to
consider the frames $e$ as well
as $e'$ for this instanton equation, as ${e}^\ast \Gamma^P$ and
${e'}^\ast \Gamma^P$ are $\mathfrak{su}(2)$-valued connection $1$-forms.
However, this raises the question whether the two
extensions $X_\mu e^\mu$ and ${X'}_\mu {e'}^\mu$ are in any sense comparable.
Unfortunately, the coframe-transformations are only required to be
$N_{\mathrm{GL}(6,\mathbb{R})}(\mathrm{SU}(2))$-valued, which implies that the
$\mathfrak{m}$-piece will, in general, not be mapped into $\mathfrak{m}$ or even
$\mathfrak{su}(3)$. Hence, one cannot simply compare both extensions, but it is
admissible to consider both cases.
\end{itemize}
In summary, these remarks were not relevant for the cases studied for
example in~\cite{Harland:2011zs,Ivanova:2012vz} or our earlier
results~\cite{Bunk:2014kva}, because the construction of the $G$-structures on
the warped product $M^5{\times_{\phi}}I$ followed immediately from the chosen
frame on
$M^5$. In other words, no (base-point dependent) transformation of coframes was
necessary. Even on our KT- and HKT-sine cones of~\cite{Bunk:2014kva}, the
relevant rescaling~\eqref{eqn:compl-structure-KT} does not affect the
computations due to conformal equivalence to the cylinder. However, here the
situation is more involved and a careful analysis is mandatory.
\subsection{Instantons on nearly Kähler sine-cones}
\subsubsection{Matrix equations - part I}
\label{subsec:NK-Instantons}
The set-up for the nearly Kähler sine-cone has been described in
Section~\ref{subsec:NK-SineCone}. In particular, we are investigating 
extensions of the connection $\Gamma^P$ on the sine-cone in this subsection.
$M^6$ being a nearly Kähler manifold, the instanton equation with respect to
the coframe $e^\mu$ is equivalent to
\begin{subequations}
\begin{alignat}{3}
 \o \wedge \o \wedge \mathcal{F} &= 0 \qquad  & &\Leftrightarrow & \qquad
\o^{\hat{\mu} \hat{\nu}} \mathcal{F}_{\hat{\mu} \hat{\nu}}&=0 \; , \\%
\Om \wedge \mathcal{F} &= 0 & &\Leftrightarrow & \Om^{\hat{\s} \hat{\mu}
\hat{\nu}} 
\mathcal{F}_{\hat{\mu} \hat{\nu}} &= 0 \qquad\text{for}\quad \hat{\s} = 1,\ldots, 6 \; .%
\end{alignat}%
\label{eqn:instanton_NK}%
\end{subequations}%
The seven equations~\eqref{eqn:instanton_NK} restrict the space of admissible
$2$-forms, and the instanton bundle, which is locally isomorphic\footnote{One
employs the identification $\mathfrak{so}(6)\simeq \La^2(\mathbb{R}^6)$ to
obtain $2$-forms from antisymmetric $6\times6$-matrices.} to the subspace
$\mathfrak{m}$, is spanned by
\begin{equation}
\begin{aligned}
 e^5 \wedge e^6 &- \sfrac{\La \sin \varphi}{4} \left( \sin \varphi \, \h_{ab}^1 
+  \cos \varphi \, \h_{ab}^3 \right) e^a \wedge e^b \and \\ %
e^a \wedge e^6 &-  \La \sin \varphi \left( \sin \varphi \, \h^{1\,a}_{\ \ \ b} 
+ \cos \varphi \, \h^{3\,a}_{\ \ \ b} \right) e^b \wedge e^5 \; .
\end{aligned}
\end{equation}
This can be seen either by direct computation or by the explicit form of the
projectors from $\mathfrak{so}(6)$ to $\mathfrak{su}(3)$ of
\cite{Ivanova:1993ws}.
Here we have used the Riemannian metric to pull up one of the indices of 
$\eta^3$, and from here on we use $e^6 = \diff r$. 

A $6$-dimensional representation of $\mathfrak{m}$
can be chosen as in~\cite{Harland:2011zs,Ivanova:2012vz},
\begin{subequations}
\begin{alignat}{2}
 (\hat{I}_5)_a^b &= \sfrac{1}{2}\, \h_{ab}^3 \; , & \quad -(\hat{I}_5)_5^6 &=
(\hat{I}_5)_6^5 = 1 \; ,\\
-(\hat{I}_a)_b^6&=(\hat{I}_a)_6^b=\de_a^b \; ,  & \quad (\hat{I}_a)_b^5 &= -
(\hat{I}_a)_5^b= \h_{ab}^3 \; ,
\end{alignat}
  \label{eqn:rep_generators}
\end{subequations}
from which one obtains the structure constants
\begin{align}
 \ddu{f}{5}{a}{b} = \sfrac{3}{2}\, \h^{3\ b}_{\ a} \und \ddu{f}{a}{b}{5} = 2\,
\h_{ab}^3 \;.%
\label{eqn:KT_structure-constants}
\end{align}
The torsion components of the canonical $\mathfrak{su}(2)$-connection 
$\Gamma^P$ in the unrotated frame $e^\mu$ read
\begin{subequations}
 \begin{align}
  T_{ab}^5 &= -2 \, \h_{ab}^3  =  -  \ddu{f}{a}{b}{5} \; , \\
 T_{b5}^a &= - \sfrac{3}{2} \, (\h^3)^a_{\ b} 
 =- \ddu{f}{b}{5}{a} \; .
 \end{align}
\end{subequations}
With the chosen representation and by inserting the ansatz
\begin{align}
 \mathcal{A} = \Gamma^P + X_\mu \, e^\mu %
\label{eqn:ansatz_NK}
\end{align}
into~\eqref{eqn:instanton_NK}, one obtains the non-vanishing components
$N_{\mu \nu}^{\rho}$ of the parametrization~\eqref{eqn:parametrise_X} as follows:
\begin{align}
 N_{ab}^5 =  \sfrac{\La \sin\varphi}{2}\, \left( \sin \varphi \, \h_{ab}^1 +  
\cos \varphi \, \h_{ab}^3 \right) \and  %
N_{b5}^a = \La \sin\varphi  \left( \sin \varphi \, \h^{1\,a}_{\ \ \ b} + \cos
\varphi \, \h^{3\,a}_{\ \ \ b} \right) \; .
\end{align}
Finally, the matrix equations for $X_\mu$ read
\begin{subequations} 
 \begin{align}
  [\hat{I}_i,X_\mu] &= \ddu{f}{i}{\mu}{\nu}\, X_\nu\;,\\
  [X_a , X_b ] &=  \frac{\La \sin\varphi}{2}\, \left( \sin \varphi \, \h_{ab}^1 
+  \cos \varphi \, \h_{ab}^3 \right)\dot{X}_5  + 2 \, \h_{ab}^3 X_5 + 
\ddu{f}{a}{b}{i} \mathcal{N}_i \; ,\\
   [X_5 , X_a] &= \La \sin\varphi \left( \sin \varphi \, \h^{1\ b}_{\ a} + \cos
 \varphi \, \h^{3\ b}_{\ a} \right) \dot{X}_b + \frac{3}{2} \, \h^{3\ b}_{\ a}
 X_b \; , 
 \end{align}
 \label{eqn:matrix-eq_NK}%
\end{subequations}
where the first line is just the equivariance condition~\eqref{eqn:equivariance}.
The dot-notation means $\dot{Y} \equiv \sfrac{\mathrm{d}}{\mathrm{d} r}Y$. 
An obvious solution to~\eqref{eqn:matrix-eq_NK} is $X_\mu\equiv0$, 
which yields the instanton solution $\Acal =\Gamma^P$ that is the lift of the 
instanton $\Gamma^P$ from $M^5$ to the sine-cone $C_s(M^5)$. 

Consider the ansatz
\begin{equation}
 X_a= \psi(r) \left(\exp(\xi \, \h^3) \right)_{a}^{\ b} \hat{I}_b \; , 
\for \xi \in [0,2 \pi) \und X_5 = \chi(r) \hat{I}_5 \; , %
\label{eqn:rotated_ansatz_NKI}
\end{equation}
which respects equivariance due to $[\h^\a , \bar{\h}^\b]=0$. Here, $\xi$ is
a parameter, and $\psi(r)$, $\chi(r)$ are two functions depending only on
the cone direction $r$. Inserting \eqref{eqn:rotated_ansatz_NKI} into 
\eqref{eqn:matrix-eq_NK} yields 
\begin{equation}
 \mathcal{N}_i = \psi^2(r) \, \hat{I}_i \, \for i = 6,7,8
\end{equation}
as well as the following differential equations
\begin{subequations}
\label{eqn:ode_NKI}
\begin{equation}
 \sfrac{\La}{2} \, \dot{\chi}(r)  \,  \sin (2 \varphi) =4 \left( \psi^2(r)- 
\chi(r) \right) \und \sfrac{\La}{2} \dot{\psi}(r) \sin (2\varphi) = 
\sfrac{3}{2} \psi(r) \left( \chi(r) -1\right) \; , 
\end{equation}
which are subject to the constraints
\begin{equation}
\sfrac{\La}{2} \, \dot{\psi}(r) \, \sin^2 \varphi = \sfrac{\La}{2} \,  
\dot{\chi}(r) \, \sin^2 \varphi  =0 \; .
\end{equation}
\end{subequations}
As a matter of fact, these equations~\eqref{eqn:ode_NKI} hold for any value of
$\xi \in [0,2 \pi)$. The solutions to~\eqref{eqn:ode_NKI} are readily obtained
to be the following:
\begin{itemize}
 \item \underline{$(\psi,\chi)=(0,0) $:}  This is, of course, the trivial
solution of \eqref{eqn:matrix-eq_NK}, but is still required for consistency as
it confirms that $\Gamma^P$ satisfies the $\Omega_s$-instanton condition on 
$M^6$.
\item \underline{$(\psi,\chi)=(1,1) $:} Here we obtain an extension of the
original instanton $\Gamma^P$. Despite being
an $\Omega_s$-instanton, this newly obtain instanton is a mere lift of an
instanton in $M^5$ as it does not have any dependence on the cone direction.
\item \underline{$(\psi,\chi)=(-1,1) $:} Again, we obtain an extension which
is, however, a lift of an $M^5$-instanton. Note that the existence of this
solutions follows from $\xi \mapsto \xi + \pi$, as  $\left(\exp(\pi\ \h^3)
\right)_{a}^{\ b} = -\delta_{a}^{\ b}$. 
\end{itemize}
Hence, we have a one-parameter family of $\mathfrak{su}(3)$-valued
instantons given by 
\begin{equation}
 A = \Gamma^P + \left(\exp(\xi \ \h^3) \right)_{a}^{\ b} \hat{I}_b \otimes
e^a + \hat{I}_5 \otimes e^5 \; .%
\label{eqn:1-para_family}
\end{equation}
To summarize, the ansatz solving the matrix equations~\eqref{eqn:matrix-eq_NK}
generates isolated instanton solutions which can all be interpreted as
lifts of connections living on $M^5$. The non-trivial solutions are
$\mathfrak{su}(3)$-valued connections; whereas the trivial solution is a purely
$\mathfrak{su}(2)$-valued connection. 

\paragraph{Remarks:} First, the family of solutions~\eqref{eqn:1-para_family} can 
be seen to be gauge orbit if we recall that $(\h^3)_{\mu}^{\nu} \propto 
\ddu{f}{5}{\mu}{\nu} =  \mathrm{ad}(I_5)_{\mu}^{\nu} $ and then $\exp(\xi \; 
\h^3) \propto \mathrm{Ad}(\exp(I_5))$. 
Nevertheless, this gauge symmetry clarifies the origin of the $\psi$-reflection 
symmetry of the solutions.

Second, in the same manner as in our previous studies~\cite{Bunk:2014kva} we 
can equivalently provide the matrix equations on the conformally equivalent 
cylinder with coordinate $\tau$ as follows:
\begin{subequations} 
 \label{eqn:matrix-eq_NK-cylinder}%
 \begin{align}
  [\hat{I}_i,X_\mu] &= \ddu{f}{i}{\mu}{\nu}\, X_\nu\;,\\
  [X_a , X_b ] &=  \frac{1}{2} \left( \sin \varphi \, \h_{ab}^1 
+  \cos \varphi \, \h_{ab}^3 \right) \frac{\diff}{\diff \tau}X_5  + 2 \, 
\h_{ab}^3 X_5 + 
\ddu{f}{a}{b}{i} \mathcal{N}_i \; ,\\
   [X_5 , X_a] &= \left( \sin \varphi \, \h^{1\ b}_{\ a} + \cos
 \varphi \, \h^{3\ b}_{\ a} \right) \frac{\diff}{\diff \tau} X_b + \frac{3}{2} 
\, \h^{3\ b}_{\ a}  X_b \; . 
 \end{align}
\end{subequations}
Further, the limit $\La \to \infty$ (with $\varphi{ =
}\sfrac{r}{\La} \to 0$ and keeping $r$ fixed) transforms the
sine-cone into the Calabi-Yau cone, as mentioned in
Subsection~\ref{subsec:NK-SineCone}. In this limit, the matrix
equations~\eqref{eqn:matrix-eq_NK-cylinder} take the following form:
\begin{equation}
 [X_a,X_b] = \ddu{f}{a}{b}{5} \left( X_5 +\frac{1}{4} \, \dot{X}_5\right) +
\ddu{f}{a}{b}{i} \mathcal{N}_i \and [X_5,X_a]= \ddu{f}{5}{a}{b}
\left( X_b + \frac{2}{3} \, \dot{X}_b \right) \;,%
\label{eqn:NK_to_CY_1}
\end{equation}
which are exactly the same equations as on the Kähler-torsion sine-cone of our
early results~\cite{Bunk:2014kva}. Applying the $\tau$-dependent version of the
ansatz~\eqref{eqn:rotated_ansatz_NKI} yields 
\begin{equation}
\dot{\chi}(\tau)= 4 \left( \psi^2(\tau) - \chi(\tau)\right) \und 
\dot{\psi}(\tau) = \sfrac{3}{2} \psi(\tau) \left(\chi(\tau) -1 \right) \; . %
\label{eqn:NK_to_CY_1a}
\end{equation}
Obviously, all constant solutions found above are still instantons
on the CY-cone, but the reduced equations do not automatically enforce constant
$\chi$ and $\psi$. Finally, note that~\eqref{eqn:NK_to_CY_1a} is, of course, 
equivalent to~\eqref{eqn:ode_NKI} in the limit $\La \to \infty$ as the 
constraint on the derivatives vanishes.

Third, the sine-cone is a conifold with two conical singularities, here at
$\varphi=0$ and $\varphi=\pi$. One observes that the coefficient functions,
i.e.  $\cos \varphi$ and $\sin \varphi$, of \eqref{eqn:matrix-eq_NK} as well as
our solutions are well-behaved at the singular points. However, recall the
remark from
Subsection~\ref{subsec:NK-SineCone} that the defining sections of the
SU$(3)$-structure become trivial at these singular points; hence, the instanton
condition is not well-defined there. Yet, in principal one could continue the 
gauge field to these points.
\subsubsection{Nearly Kähler canonical connection}
\label{subsec:CanonConSU(3)}
In this section we construct the canonical $\mathfrak{su}(3)$-connection
of the nearly Kähler sine-cone.
It turns out that we obtain an instanton for the SU$(3)$-structure
that is not the lift of an instanton on $M^5$; furthermore, this instanton is 
of the form~\eqref{eqn:ansatz_NK} presented above.
On the $5$-manifold $M^5$ the Maurer-Cartan equations read
\begin{subequations}
 \begin{align}
  \diff e^a &= - {(\Gamma^P)}_b^a \wedge e^b + \sfrac{1}{2}\, T_{\mu 
\nu}^a\,e^\mu \wedge
e^\nu \; , \\
\diff e^5 &= -{(\Gamma^P)}_5^5 \wedge e^5 + \sfrac{1}{2}\, T_{\mu \nu}^5\,e^\mu 
\wedge e^\nu
\; ,
 \end{align}
  \label{eqn:Maurer_Cartan}
\end{subequations}
where the torsion components are given by (cf.~\cite{Harland:2011zs,Ivanova:2012vz})
\begin{align}
 T_{b5}^a = -\sfrac{3}{2}\, \h^{3a}_{\ \ b} \und T_{ab}^5 = -2\, \h_{ab}^3\; .
\end{align}
In particular, the last identity implies ${(\Gamma^P)}_5^5=0$ due to the 
Sasaki-Einstein relation $\diff e^5 = - 2\, \o^3$.

Next, we are interested in the Maurer-Cartan equations for the frame $e_s^\mu$
resulting from the rotation~\eqref{eqn:rotation_SU(2)-structure} and 
rescaling~\eqref{eqn:SU(3)_nearlKähler} of the SU$(2)$-structure. With respect 
to coframes $e$ adapted to $\mathcal{Q}$, the canonical
$\mathfrak{su}(2)$-connection $\Gamma^P$ has components
\begin{align}
{(\Gamma^P)}_\mu^\nu = {(\Gamma^P)}^i \ddu{f}{i}{\mu}{\nu} 
\qquad\text{with}\quad
(\ddu{f}{i}{a}{b}) \propto \bar{\h}^{\a(i)}\:,
\end{align}
where $\a(i) = i{-}5$ and $\bar{\h}^\a$ are the anti-self-dual 't~Hooft tensors.
Noting that $[ \h^\a , \bar{\h}^\b]=0$ for all $\a, \b$, we see that the components of the canonical $\mathfrak{su}(2)$-connection are unaffected by the
homogeneous part of the transformation~\eqref{eqn:trafo_connection} with
\begin{equation}
\label{eqn:trafo_cy_nksc}
 L(r) = \Lambda \sin(\varphi)\, \begin{pmatrix} \exp(\frac{\varphi}{2} \, 
\h^2)_{4\times4} & 0_{4\times2}\\ 0_{2\times4} & 
\mathbbm{1}_{2\times2}  \end{pmatrix} \in 
N_{\mathrm{GL}(6,\mathbb{R})}(\mathrm{SU}(2))\;,
\end{equation}
which realizes the
rotation~\eqref{eqn:rotation_SU(2)-structure} and the
rescaling~\eqref{eqn:SU(3)_nearlKähler}.
In detail, the transformation reads ${(\Gamma^P)}^a_b = L^a_c\, 
{(\Gamma^P)}^c_d\, (L^{-1})^d_b$.
A straightforward computation yields
\begin{subequations}
\begin{align}
 \diff e_s^a &= -{(\Gamma^P)}_b^a \wedge e_s^b - \frac{\cot\varphi}{\La} \left( 
e_s^a \wedge e_s^6 + \h^{3a}_{\ \ b}\, e_s^b \wedge e_s^5 \right) - 
\frac{\cot\varphi}{2 \La} \h^{3a}_{\ \ b}\, e_s^b \wedge e_s^5 \\*
 &\phantom{= -{(\Gamma^P)}_b^a \wedge e_s^b \; }-\frac{1}{2 \La} \left( 
\h^{2a}_{\ \ 
b}\, e_s^b \wedge e_s^6 - \h^{1a}_{\ \ b}\, e_s^b \wedge e _s^5\right) + 
\frac{1}{\La} \h^{1a}_{\ \ b}\, e_s^b \wedge e_s^5  \; , \notag\\
 \diff e_s^5 &= -\frac{\cot\varphi}{\La} \left(e_s^5 \wedge e_s^6 + \h_{ab}^3\, e_s^a \wedge e_s^b \right) + \frac{1}{\La} \h_{ab}^1\, e_s^a \wedge e_s^b \; ,\\
\diff e_s^6 &=0 \; .
\end{align}%
\label{eqn:rotated_Maurer_Cartan}%
\end{subequations}%
It is important to realize that, although the components ${(\Gamma^P)}^a_b$ 
used  in~\eqref{eqn:rotated_Maurer_Cartan} coincide with the components of the 
lift of  the canonical connection on the Sasaki-Einstein $5$-manifold to the 
cylinder, the transformed coframe $e_s^\mu$ is used since we are on the nearly 
Kähler sine-cone.
Thus, ${(\Gamma^P)}^a_b$ no longer comprises the canonical 
$\mathfrak{su}(2)$-connection; however, it forms a different 
$\mathfrak{su}(2)$-valued connection $\Gamma_{\mathfrak{su}(2)}$.
This is because the inhomogeneous term in~\eqref{eqn:trafo_connection}, which
results from the change of basis, has been split off.

Introducing an almost complex structure $J$ via demanding
\begin{align}
 \Th_s^1= e_s^1 + \im e_s^2  \; , \quad \Th_s^2= e_s^3 + \im e_s^4  \; , \quad
\Th_s^3= \im(e_s^5 + \im e_s^6) \; %
\label{eqn:complex_basis}
\end{align}
to be $(1,0)$-forms yields
\begin{equation}
 \diff {\begin{pmatrix}  \Th_s^1 \\ \Th_s^2 \\ \Th_s^3  \end{pmatrix}} = -
\underbrace{%
 \begin{pmatrix} %
  {\hat{\Gamma}_{\mathfrak{su}(2)}}{}_1^1 + \frac{\im \cot \varphi}{2 \La} e_s^5 & {\hat{\Gamma}_{\mathfrak{su}(2)}}{}_2^1 & -\frac{\cot \varphi}{\La} \Th_s^1 - \frac{1}{2 \La} \Th_s^{\bar{2}} \\ %
  {\hat{\Gamma}_{\mathfrak{su}(2)}}{}_1^2 & {\hat{\Gamma}_{\mathfrak{su}(2)}}{}_2^2 + \frac{\im \cot \varphi}{2 \La} e_s^5 & -\frac{\cot \varphi}{\La} \Th_s^2 + \frac{1}{2 \La} \Th_s^{\bar{1}} \\ %
 \frac{\cot \varphi}{\La} \Th_s^{\bar{1}} + \frac{1}{2 \La} \Th_s^2 & \frac{\cot\varphi}{\La} \Th_s^{\bar{2}} - \frac{1}{2 \La} \Th_s^1 & -\frac{\im \cot\varphi}{\La} e_s^5 %
 \end{pmatrix}%
}_{\text{canonical }\mathfrak{su}(3)\text{-connection }
\hat{\Gamma}_{\mathfrak{su}(3)} \text{ on sine-cone}  }%
\wedge \begin{pmatrix}  \Th_s^1 \\ \Th_s^2 \\ \Th_s^3  \end{pmatrix} %
\underbrace{%
 - \frac{1}{\La} \begin{pmatrix}  \Th_s^{\bar{2}\bar{3}} \\ \Th_s^{\bar{3}\bar{1}} \\ \Th_s^{\bar{1}\bar{2}} \end{pmatrix}%
}_{\text{NK-torsion } \hat{T}}.%
\label{eqn:preliminary}
\end{equation}
Here we used the shorthand notation $\Theta^{\bar{\a}\bar{\b}} \equiv
\Theta^{\bar{\a}} \wedge \Theta^{\bar{\b}}$.

The connection $1$-forms ${\hat{\Gamma}_{\mathfrak{su}(2)}}{}_{\a}^{\b}$ with 
$\a,\b = 1,2$ are defined via the components ${(\Gamma^P)}_{a}^{b}$ by employing 
\eqref{eqn:Maurer_Cartan} and \eqref{eqn:rotated_Maurer_Cartan} as well as the 
change to the complex basis~\eqref{eqn:complex_basis}.
We use the hat to indicate that we are considering the connection forms with respect to the complex basis $\Theta_s$ rather than the real basis $e_s$.
Thus, the corresponding Maurer-Cartan equations read
\begin{equation}
 \diff \Th_{s}^{\a} = - {\hat{\Gamma}_{\mathfrak{su}(3)}}{}_{\b}^{\a} \wedge \Th_{s}^{\b} + \hat{T}^{\a} \und \diff
\Th_{s}^{\bar{\a}} = - {\hat{\Gamma}_{\mathfrak{su}(3)}}{}_{\bar{\b}}^{\bar{\a}} \wedge
\Th_{s}^{\bar{\b}} + \hat{T}^{\bar{\a}} \; .
\label{eqn:MCSU(3)}
\end{equation}
Note that $\Gamma_{\mathfrak{su}(3)}
=\text{diag}\bigl(\hat{\Gamma}_{\mathfrak{su}(3)}^{\phantom{*}},\hat{\Gamma}^*_{
\mathfrak { su }(3)}\bigr)$ 
is indeed a connection on
$TM^6$, which can be seen from~\eqref{eqn:MCSU(3)} and the fact that $\hat{T}$
transforms as a tensor.
Furthermore, $\Gamma_{\mathfrak{su}(3)}$ is an instanton because it
satisfies the conditions of proposition 3.1 of~\cite{Harland:2011zs}.

The above result~\eqref{eqn:preliminary} can be brought into a more suggestive
form by rewriting it as
\begin{subequations}
\label{eqn:CanonConSU3}
\begin{align}
 \hat{\Gamma}_{\mathfrak{su}(3)} =  \hat{\Gamma}_{\mathfrak{su}(2)} &+ %
\frac{1}{2 \Lambda} \begin{pmatrix} 0 & 0 & -2 \cot\varphi \\ %
0 & 0 & 1 \\ %
2 \cot\varphi & - 1 & 0 %
\end{pmatrix}%
e_s^1 + %
\frac{\im}{2 \Lambda} \begin{pmatrix} 0 & 0 & -2 \cot\varphi\\%
0 & 0 & -1 \\%
- 2 \cot\varphi & -1 & 0%
\end{pmatrix}%
e_s^2 \notag \\*
&+ \frac{1}{2 \Lambda} \begin{pmatrix} 0 & 0 & -1 \\ %
0 & 0 & - 2 \cot\varphi \\ %
1 & 2 \cot\varphi & 0 %
\end{pmatrix}%
e_s^3  %
+ \frac{\im}{2 \Lambda} \begin{pmatrix} 0 & 0 & 1 \\ %
0 & 0 & - 2 \cot\varphi \\ %
1 &  -2\cot\varphi & 0 %
\end{pmatrix}%
e_s^4 \\*
&+ \frac{\im}{2 \Lambda} %
\begin{pmatrix} \cot\varphi & 0 & 0 \\ %
0 & \cot\varphi & 0 \\ %
0 & 0 & -2\cot\varphi %
\end{pmatrix} %
e_s^5 \notag \\%
= \hat{\Gamma}_{\mathfrak{su}(2)} &+ B_\mu \otimes e_s^\mu\; ,%
\end{align}
\end{subequations}
which reflects exactly the $X_\mu$-ansatz from \eqref{eqn:ansatz_NK}.
One can check that the matrices $B_\mu$ satisfy the equivariance condition~\eqref{eqn:equivariance}.
Thus, as $\Gamma_{\mathfrak{su}(3)}$ is a connection on $TM^6$, one can infer by
the same arguments as in Section~\ref{sec:InsttoMatrix} that
$\Gamma_{\mathfrak{su}(2)}$ is a well-defined connection on $TM^6$.
An alternative way to see that is to check that the inhomogeneous part, 
which has been split
off in the transformation law~\eqref{eqn:trafo_connection} for the components of
$\Gamma^P$, glues to globally well-defined $1$-forms with values in the adjoint 
bundle of $\mathcal{P}$. This, however, holds due to the fact that the 
transformation $L$ given in~\eqref{eqn:trafo_cy_nksc} commutes with the 
SU$(2)$ subgroup of GL$(6,\mathbb{R})$, i.e. takes values in centralizer 
$C_{\mathrm{GL}(6,\mathbb{R})}(\mathrm{SU}(2))$.

Note that in the limit $\La \to \infty$ (i.e.~$\varphi = \frac{r}{\Lambda} \to 0$) the torsion on $C(M^5)$
vanishes, and $\hat{\Gamma}_{\mathfrak{su}(3)}$ coincides with the 
connection corresponding to the $\chi=\psi=1$ case of \cite{Harland:2011zs}, 
which has been stated to be the Levi-Civita connection of the cone. 
Furthermore, this is consistent with the observation that as 
$\hat{\Gamma}_{\mathfrak{su}(3)}$ preserves the metric and as  in the above 
limit its torsion vanishes, $\hat{\Gamma}_{\mathfrak{su}(3)}$ has to 
converge to the Levi-Civita connection of the CY-cone.
\subsubsection{Matrix equations - part II}
\label{subsec:NK-InstantonsII}
As pointed out above, there are two different $\mathfrak{su}(2)$-valued connections on
the nearly Kähler sine-cone.
On the one hand, there is the lift of the canonical connection $\Gamma^P$ of the
Sasaki-Einstein $5$-manifold; on the other hand, there is
$\Gamma_{\mathfrak{su}(2)}$.
Remarkably, the respective curvature 2-forms coincide, i.e.
\begin{equation}
 R_{\Gamma^P} = R_{\Gamma_{\mathfrak{su}(2)}}\;.
\end{equation}
This stems from the fact that the generators of the two
transformations~\eqref{eqn:rotation_SU(2)-structure} and
\eqref{eqn:SU(3)_nearlKähler}, which lead from the cylinder to the sine-cone,
commute with $\mathfrak{su}(2)$. In other words, the inhomogeneous part
of~\eqref{eqn:trafo_connection} yields an abelian flat part proportional to
$e_s^6$.
As a consequence, $\Gamma_{\mathfrak{su}(2)}$ is another
$\mathfrak{su}(2)$-valued instanton on the sine-cone, since $\Gamma^P$ is an
instanton itself\,\footnote{Recall Subsection~\ref{subsec:remarks_Instanton}: 
$\Gamma^P$ is a connection on the SU$(2)$-bundle $\mathcal{Q}$, whereas 
$\Gamma_{\mathfrak{su}(2)}$ is a connection on the SU$(2)$-bundle 
$\mathcal{Q}'$.}.
Therefore, we can use $\Gamma_{\mathfrak{su}(2)}$ in the procedure described in
Section~\ref{sec:InsttoMatrix}: One extends $\Gamma_{\mathfrak{su}(2)}$
by some suitable $1$-form $X_\mu \, e_s^\mu$ and investigates the conditions on
$X_\mu$ such that the new connection is an instanton on the
sine-cone.

However, we have to adjust the equations~\eqref{eqn:matrix-eq_NK} due
to the different torsion of $\Gamma_{\mathfrak{su}(2)}$.
Denoting by $T$ the torsion of $\Gamma^P$, the torsion of 
$\Gamma_{\mathfrak{su}(2)}$ reads
\begin{equation}
 T_{\mathfrak{su}(2)}^{\hat{\mu}} = T^{\hat{\mu}} + \frac{1}{\Lambda} \Big( 
\delta^{\hat{\mu}}_{\hat{\nu}}\,\cot\varphi + \sfrac{1}{2}\, 
\eta^{2\hat{\mu}}_{\ \ \hat{\nu}} \Big)\, e_s^6 \wedge e_s^{\hat{\nu}}\;, %
\label{eqn:torsion_gamma_su2}
\end{equation}
where we defined $\eta^{2}_{\hat{\mu} \hat{\nu}} = \eta^{2}_{a b}$
 for $\hat{\mu},\hat{\nu} =a,b\in \lbrace 1,\ldots,4 \rbrace$ and
$\eta^{2}_{\hat{\mu} \hat{\nu}} = 0$ whenever $\hat{\mu} \geq 5$ or
$\hat{\nu} \geq 5$.
The components of $N$ are the same as in
Subsection~\ref{subsec:NK-Instantons} and, by inserting everything
into~\eqref{eqn:F(canonical)withN}, we obtain the matrix equations
\begin{subequations} 
\begin{align}
[\hat{I}_i,X_\mu] &= \ddu{f}{i}{\mu}{\nu}\, X_\nu \;,\\ %
  [X_a , X_b ] &= \frac{1}{2}\, \eta^3_{ab} \dot{X}_5 + \frac{1}{2 \Lambda}
 \left(  5 \cot\varphi  \, \h^3_{ab} - 4 \, \h_{ab}^1 \right) \, X_5
 +  \ddu{f}{a}{b}{i}\, \mathcal{N}_i \; ,\\ %
  [X_5 , X_a] &= \eta^{3\ b}_{\ a} \dot{X}_b + \frac{1}{2 \La}\, \left( 5\,
\cot\varphi \, \eta^{3\ b}_{\ a}  - 3\, \h^{1\ b}_{\  a} - 
 \eta^{3\ c}_{\ a}\, \eta^{2 \ b}_{\ c}  \right) \,X_b
\; ,
\end{align}
  \label{eqn:matrix-eq_NKII}%
\end{subequations}
with the notation $\dot{Y}=\sfrac{\mathrm{d} }{\mathrm{d} r} Y$.
Next, we use the matrices in~\eqref{eqn:CanonConSU3} for the extension of
$\Gamma_{\mathfrak{su}(2)}$.
Recall that we had defined auxiliary matrices $B_\mu$ that solve the
equivariance condition~\eqref{eqn:equivariance} by
writing~\eqref{eqn:CanonConSU3} in the form
\begin{equation}
 \hat{\Gamma}_{\mathfrak{su}(3)} = \hat{\Gamma}_{\mathfrak{su}(2)} + B_\mu
e^\mu_s \; ,
\end{equation}
and that the $B_\mu$ explicitly depend on $\varphi = \sfrac{r}{\La}$.
Hence, we may set 
\begin{equation}
\label{eqn:ansatz_NKII}
 X_a \coloneqq \psi(r)\, B_a \und X_5 \coloneqq \chi(r)\, B_5
\end{equation}
as in the usual procedure\footnote{Note that in~\eqref{eqn:CanonConSU3} we have 
$B_\mu 
\in \mathrm{End}(\mathbb{C}^3)$. Here we used the identification 
$\mathbb{C} \simeq \mathbb{R}^2$ to obtain $B_\mu \in 
\mathrm{End}(\mathbb{R}^6)$, which is necessary for the 
ansatz~\eqref{eqn:ansatz_generic}.}.
The equivariance condition enforces the same coefficient function
$\psi(r)$  for all four $B_a$.
Inserting this $X_\mu$-ansatz in the matrix
equations~\eqref{eqn:matrix-eq_NKII}, one can first of all read off 
 \begin{equation}
  \mathcal{N}_i = \psi(r)^2 \frac{1+4 \cot^2\!\varphi}{4 \La^2} I_i \,, 
\for i = 6,7,8 \;,
 \end{equation}
which is compatible with the assumptions on $\mathcal{N}$ used in 
Subsection~\ref{sec:InsttoMatrix}.
Using this explicit form, we obtain the algebraic equation
\begin{equation}
\psi(r)^2 - \chi(r) = 0\;.
 \label{eqn:algebraiceom}
\end{equation}
This then reduces the remaining equations to
\begin{equation}
 \dot{\chi}(r) = \dot{\psi}(r) = 0 \und  \psi(r)
\big(\chi(r) - 1\big) = 0\;.
\end{equation}
Let us now comment on the three solutions to this system:
\begin{itemize}
 \item \underline{$(\psi,\chi) = (0,0)$:} To start with, there is the obvious
trivial solution of \eqref{eqn:matrix-eq_NKII}. This is required for
consistency, since $\Gamma_{\mathfrak{su}(2)}$
is an instanton.
 \item \underline{$(\psi,\chi) = (1,1)$:} This second solution is very
important because it reproduces $\Gamma_{\mathfrak{su}(3)}$ from
Subsection~\ref{subsec:CanonConSU(3)}.
We already knew from proposition 3.1 of~\cite{Harland:2011zs} that this
particular connection is an instanton on the nearly Kähler sine-cone, but here
we confirmed it directly, using techniques completely different than those
employed in~\cite{Harland:2011zs}.
In addition, this provides us with another way of constructing the canonical
connection of the nearly Kähler sine-cone than the one we followed in
Subsection~\ref{subsec:CanonConSU(3)}, namely as the extension of an
$\mathfrak{su}(2)$-valued instanton.
 \item \underline{$(\psi,\chi) = (-1,1)$:} Third, there is again the solution
which results from the invariance of \eqref{eqn:matrix-eq_NKII} under the
simultaneous sign-flip $X_a \mapsto -X_a$ for $a=1,2,3,4$. Nevertheless, this
solution is an additional instanton.
\end{itemize}
In summary, the solutions we obtained here are isolated $\mathfrak{su}(3)$- and
$\mathfrak{su}(2)$-valued connections on $M^6$ that cannot be traced back to
lifts of connections on $M^5$.
In contrast to e.g.~\cite{Gemmer:2011cp}, there are no instanton solutions that
interpolate between these isolated instantons.
\paragraph{Remarks:} 
First, the CY-limit $\La \to \infty$ of~\eqref{eqn:matrix-eq_NKII} is given by
\begin{equation}
[X_a,X_b]= \ddu{f}{a}{b}{5} \left(X_5 + \frac{1}{4} \,
\frac{\diff}{\diff \tau}X_5 \right) + \ddu{f}{a}{b}{i} \mathcal{N}_i \and
[X_5,X_a]= \ddu{f}{5}{a}{b} \left( X_b + \frac{2}{3} \,
\frac{\diff}{\diff \tau} X_b \right) \; ,
\end{equation}
wherein one requires the rescaling $X_\mu \mapsto \sfrac{1}{r} X_\mu$, which
can be seen from $X_\mu e_s^\mu \to X_\mu \, r e^\mu$  for $\La \to
\infty$. Further, recall that in the limit $\La \to \infty$ we have $\diff \tau
= \sfrac{1}{r} \diff r$. The above matrix equations coincide with the ones
obtained in the Kähler-torsion case of~\cite{Bunk:2014kva} as well as with the
limit~\eqref{eqn:NK_to_CY_1} of Subsection~\ref{subsec:NK-Instantons}.
Remarkably, the two reductions of Subsections~\ref{subsec:NK-Instantons} and
\ref{subsec:NK-InstantonsII} used the different $\mathfrak{su}(2)$-instantons
$\Gamma^P$ and $\Gamma_{\mathfrak{su}(2)}$ as starting point; however, in the
above limit the difference
\begin{equation}
 \Gamma^P - \Gamma_{\mathfrak{su}(2)} \xrightarrow{\La \to \infty}
\mathbbm{1} \otimes \frac{\diff r}{r} \in \Om^1(M^6,\mathrm{End}(\mathbb{R}^6))
\end{equation}
becomes an abelian flat part, which contributes to the instanton
equation via the altered torsion. 

Second, note the explicit impact of the conical singularities at $\varphi=0$ or
$\varphi=\pi$ in the matrix equations~\eqref{eqn:matrix-eq_NKII} as well as the
$B_\mu$-matrices of~\eqref{eqn:CanonConSU3}. However, we do not have to
consider these singularities, as the instanton equation is not well-defined at the tips.
\subsubsection{Transfer of solutions}
The previous subsections considered the nearly Kähler sine-cone from two 
perspectives: in Subsection~\ref{subsec:NK-Instantons} we extended the 
instanton $\Gamma^P$, which is a connection on $\mathcal{Q}$; whereas,
Subsection~\ref{subsec:NK-InstantonsII} was concerned with 
$\Gamma_{\mathfrak{su}(2)}$, being an $\mathfrak{su}(2)$-valued connection on 
$\mathcal{Q}'$, as a starting point for our ansatz~\eqref{eqn:ansatz_generic}. 
The local representations of these are related via a transformation $L$ as 
considered in~\eqref{eqn:trafo_cy_nksc}. Due to the properties of $L$ we arrive 
at the following statement (c.f.~Subsection~\ref{subsec:remarks_Instanton}):
\begin{align}
 {e'}^\ast \Gamma_{\mathfrak{su}(2)} = {e'}^\ast \Gamma^P - L^{-1} \diff L = 
{e}^\ast \Gamma^P \; ,
\end{align}
implying that $\Gamma_{\mathfrak{su}(2)}$ and $\Gamma^P$ have the same 
components with respect to their adapted coframes $e'$ and $e$. Observe that 
the inhomogeneous part that is split off in the connection $1$-form enters 
in the torsion~\eqref{eqn:torsion_gamma_su2} of $\Gamma_{\mathfrak{su}(2)}$, 
thus altering the matrix equations. However, from~\eqref{eqn:F(canonical)withN} 
one can check that the local expressions of the respective field strengths of 
the extension of both $\Gamma^P$ and $\Gamma_{\mathfrak{su}(2)}$ by $X_\mu 
\otimes {e'}^\mu =  X_\mu L_{\nu}^{\mu} \otimes e^\nu$ coincide. Consequently, 
every instanton extension $X_\mu $ of $\Gamma_{\mathfrak{su}(2)}$ gives rise to 
an instanton extension $ X_\nu L_{\mu}^{\nu}$ of $\Gamma^P$ and vice versa. In 
other words, we have the relation
\begin{equation}
\label{eqn:relation_X_mu}
  X_\mu \text{ solves \eqref{eqn:matrix-eq_NKII}} \quad \xLeftrightarrow{\quad 
1:1 \quad}  \quad X_\nu L_\mu^\nu \text{ solves \eqref{eqn:matrix-eq_NK}} \; .
\end{equation}
As a remark, the above is true if and only if $L$ takes values in the 
centralizer $C_{\mathrm{GL}(6,\mathbb{R})}(\mathrm{SU}(2))$, as then $L^{-1} 
\diff L$ gives rise to a well-defined equivariant $1$-form.

However, one should not naively expect that the solutions obtained in 
Subsections~\ref{subsec:NK-Instantons} and \eqref{subsec:NK-InstantonsII} are 
related via~\eqref{eqn:relation_X_mu}, as this does not necessarily transform 
the employed ansätze into one another. 

The benefit from observation~\eqref{eqn:relation_X_mu} is that we can generate 
further instanton solutions from our previous ones.

On the one hand, we can apply the above to~\eqref{eqn:1-para_family} and obtain 
the ansatz
\begin{equation}
 X_a = \frac{\psi(r)}{\La \sin (\sfrac{r}{\La})} \, \left( \exp\left(\frac{r}{2 
\La} \h^2\right) \, \exp(\xi \h^3) \right)_{a}^{\ b} \hat{I}_b \and X_5 = 
\frac{\chi(r)}{\La \sin (\sfrac{r}{\La})} \hat{I}_5 \; ,
\end{equation}
which inserted into~\eqref{eqn:matrix-eq_NKII} has precisely the 
solutions $(\psi,\chi)= (0,0) , (\pm 1 , 1)$, just as one would expect 
from the above arguments. This is another non-constant instanton extension for 
$\Gamma_{\mathfrak{su}(2)}$.

On the other hand, the same can be done for~\eqref{eqn:ansatz_NKII} in the 
other direction. There one derives the ansatz
\begin{equation}
\label{eqn:ansatz_NKI.1}
 X_a= \psi(r)\, \La \sin (\sfrac{r}{\La})  \exp\left(-\frac{r}{2 
\La} \h^2\right)_{a}^{\ b} B_b(r) \and X_5 = 
\chi(r) \, \La \sin (\sfrac{r}{\La})  B_5(r) \; .
\end{equation}
Rewritten in a linear combination of the $\hat{I}_\mu$, the 
ansatz~\eqref{eqn:ansatz_NKI.1} is given as
\begin{equation}
\label{eqn:ansatz_NKI.2}
\begin{aligned}
 X_1 &= \psi(r) \, \left( \cos^3\left(\sfrac{r}{2 \La}\right) \hat{I}_1 - 
\sin^3\left(\sfrac{r}{2 \La}\right) \hat{I}_3 \right) \; , \\ 
 X_2 &= \psi(r) \, \left( \cos^3\left(\sfrac{r}{2 \La}\right) \hat{I}_2 + 
\sin^3\left(\sfrac{r}{2 \La}\right) \hat{I}_4 \right) \; , \\ 
 X_3 &= \psi(r) \, \left( \cos^3\left(\sfrac{r}{2 \La}\right) \hat{I}_3 + 
\sin^3\left(\sfrac{r}{2 \La}\right) \hat{I}_1 \right) \; , \\ 
 X_4 &= \psi(r) \, \left( \cos^3\left(\sfrac{r}{2 \La}\right) \hat{I}_4 - 
\sin^3\left(\sfrac{r}{2 \La}\right) \hat{I}_2 \right) \; , \\ 
X_5 &=  \chi(r) \, \cos \left(\sfrac{r}{\La}\right)  \hat{I}_5 \; .
\end{aligned}
\end{equation}
One can check that this, again, produces the solutions $(\psi,\chi)= (0,0) , 
(\pm 1 , 1)$. Remarkably, the two non-trivial instanton solutions correspond to 
non-constant extensions of $\Gamma^P$. 

\subsection{Instantons on half-flat cylinders}
\label{subsec:HF_instantons}
Let us now return to the half-flat $6$-manifolds constructed in
Section~\ref{subsec:HF-Cylinder} and apply the ansatz developed above to
the instanton equation on these spaces.
The instanton equation on spaces with non-vanishing $W_2$ was introduced in
\eqref{eqn:def_instanton}.
In a local coframe adapted to the SU$(3)$-structure imposing the pseudo-holomorphicity condition
\begin{equation}
 \Omega_z \wedge \mathcal{F} = 0 \label{eqn:HF_instanton_1}
\end{equation}
yields the set of six equations, precisely as it has been in the nearly Kähler
case.
But the additional equation implied by the pseudo-holomorphicity condition reads
\begin{equation}
 \diff \Omega_z \wedge \mathcal{F} = 0 \qquad \Leftrightarrow \qquad
\mathcal{F}_{12} + \mathcal{F}_{34} + \sfrac{4}{3}\varrho^2\,\mathcal{F}_{56} =
0\ \label{eqn:HF_instanton_2}
\end{equation}
in the rotated frame $e_z$. Note that for $\varrho = \pm \frac{\sqrt{3}}{2}$
this coincides with the nearly Kähler instanton equation of
Subsection~\ref{subsec:NK-Instantons}, although the SU$(3)$-structure is not nearly
Kähler (see for example the torsion classes~\eqref{eqn:torsion_hfcylinder}).

It is important to recall that the lift of the canonical connection of the
Sasaki-Einstein $M^5$ provides an instanton on the cylinder that one can extend
by some $X$ in our ansatz to $\mathfrak{su}(3)$-valued connections, being 
defined either on $\mathcal{P}$ or $\mathcal{P}'$. We will do so in two 
set-ups: first, we formulate the
matrix equations in the frame $e^\mu$ and, second, the analogous computation is
performed in the adapted frame $e_z^\mu$ for the half-flat $SU(3)$-structure.
\subsubsection{Matrix equations - part I}
In the unrotated frame $e^\mu$ the instanton bundle is locally spanned by 
\begin{equation}
 e^5 \wedge e^6 - \sfrac{\varrho}{3} \left( \cos \zeta \, \h_{ab}^1 - \sin
\zeta \, \h_{ab}^2 \right) e^a \wedge e^b \and e^a  \wedge e^6 - \varrho \left(
\cos \zeta \, \h_{\ \ b}^{1 a} - \sin \zeta \, 
\h_{\ \ b}^{2 a}  \right)
e^b \wedge e^5 \;,
\end{equation}
from which we can extract the components of $(N_{\mu \nu}^\rho)$ to be
\begin{equation}
 N_{ab}^5 = \frac{2 \varrho}{3} \left( \cos \zeta \, \h_{ab}^1 - \sin
\zeta \, \h_{ab}^2   \right) \and N_{b5}^a = \varrho \left(
\cos \zeta \, \h_{\ \ b}^{1 a} - \sin \zeta \, 
\h_{\ \ b}^{2 a}  \right) \; .
\end{equation}
As the torsion components are unchanged we can directly formulate the matrix
equations
\begin{subequations}
 \begin{align}
  [\hat{I}_i,X_\mu] &= \ddu{f}{i}{\mu}{\nu}\, X_\nu\;,\\
 [X_a , X_b ] &= \frac{2 \varrho}{3} \left( \cos \zeta \, \h_{ab}^1 - \sin
\zeta \, \h_{ab}^2   \right) \dot{X}_5 + 2\, \h_{ab}^3\,
X_5 + \ddu{f}{a}{b}{i}\,
\mathcal{N}_i \; ,\\ 
 [X_5 , X_a] &= \varrho \left(
\cos \zeta \, \h_{\ a}^{1 \ b} - \sin \zeta \, 
\h_{\ a}^{2 \ b}  \right) \dot{X}_b  + \frac{3}{2}\, \h_{\ a}^{3 \ b}\, X_b\; .
 \end{align}
\label{eqn:inst-eqns_HFI}
\end{subequations}
The ansatz
\begin{equation}
 X_a= \psi(r) \left(\exp(\xi \ \h^3) \right)_{a}^{\ b} \hat{I}_b \; \for
\xi \in [0,2 \pi)  \and X_5 = \chi(r) \hat{I}_5
\end{equation}
satisfies, again, the equivariance condition of
\eqref{eqn:inst-eqns_HFI} and we obtain 
\begin{equation}
 \mathcal{N}_i = \psi^2 (r) \; \hat{I}_i \, , \for i = 6,7,8
\end{equation}
as well as the set of equations 
\begin{equation}
  \dot{\psi}(r)=\dot{\chi}(r)=0 \, , \qquad \psi^2(r) = \chi(r) \, , \und
\psi(r) \left( \chi(r) -1\right)=0 \,.
\end{equation}
for the two functions $\psi$ and $\chi$, and the equations hold for all
values of $\xi.$ Interestingly, the solutions to these
equations are identical to the nearly Kähler case~\eqref{eqn:ode_NKI}
\begin{itemize}
 \item \underline{$(\psi,\chi)=(0,0) $:} The trivial solution appears again for
consistency.
  \item \underline{$(\psi,\chi)=(\pm 1,1) $:} These two extensions of the lift
of $\Gamma^P$ are newly obtained $\Omega_z$-instantons; however, they correspond
to lifts of $M^5$-instantons because they are independent of the cylinder
direction. Recall that $(\psi,\chi)=(- 1,1)$ can be generated  from
$(\psi,\chi)=(+1,1) $ by the shift $\xi \mapsto \xi + \pi$.
\end{itemize}
Identically to the nearly Kähler case, one obtains the
one-parameter family~\eqref{eqn:1-para_family} as a solution. 

As a matter of fact, these instanton solutions are identical to the ones
obtained in Subsection~\ref{subsec:NK-Instantons}. The explanation is as
follows: first, note that nearly Kähler $6$-manifolds are a subset of
half-flat $6$-manifolds; thus, any nearly Kähler instanton solution must
necessarily appear in the half-flat scenario. Second, the matrix
equations~\eqref{eqn:matrix-eq_NK} and \eqref{eqn:inst-eqns_HFI} differ only in
their derivative parts, i.e. in the coefficients of $\dot{X}_\mu$, which
implies that both sets have coinciding constant solutions. 
\subsubsection{Matrix equations - part II}
Contrary to the previous subsection, here the focus is on the formulation of
the instanton equations in the adapted coframe $e_z^\mu$ for the
SU$(3)$-structure on the cylinder. As with respect to these, the 
SU$(3)$-structure forms have their
standard components, one only has to compute the components of its torsion with
respect to the transformed basis.

The space $\mathfrak{m}$ is now spanned by the $2$-forms
\begin{align}
 e_z^5 \wedge e_z^6 - \sfrac{1}{3}\varrho^2\, \h_{ab}^3\, e_z^a \wedge e_z^b
 \und e_z^a \wedge e_z^6 - \h^{3a}_{\ \ b}\, e_z^b \wedge e_z^5 \; ,
\end{align}
which follows from direct evaluation of \eqref{eqn:HF_instanton_1} and
\eqref{eqn:HF_instanton_2}. 
In the coframe $e_z$ the torsion components of the lifted canonical connection of the Sasaki-Einstein manifold are
\begin{equation}
 {\tilde{T}^5}_{ab} = 2 \varrho \, \eta^1_{ab} \und \tilde{T}_{b5}^a
= \sfrac{3}{2 \varrho}\,\eta^{1a}_{\ \ b}\;.
\end{equation}
In addition, we need the tensor $N$ that appeared in
\eqref{eqn:parametrise_X}.
Since the instanton equations take a slightly different form here, its components
now read
\begin{equation}
 N^a_{\mu \nu} = \sfrac{2}{3}\, {f_{\mu \nu}}^a \und N^5_{\mu \nu} = \sfrac{1}{3}\varrho^2\, {f_{\mu \nu}}^5\;,
\end{equation}
wherein we have used the same $\mathfrak{su}(3)$ structure constants as in~\eqref{eqn:KT_structure-constants}.
With these alterations \eqref{eqn:parametrise_X} can be written as
\begin{subequations}
 \begin{align}
 [\hat{I}_i,X_\mu] ={}& \ddu{f}{i}{\mu}{\nu}\, X_\nu\;,\\
  [X_a,X_b] ={}& -2 \varrho \, \eta^1_{ab}\, X_5 + \sfrac{2}{3}\varrho^2\,
{f_{ab}}^5\, \dot{X}_5 + \mathcal{N}_i\, {f_{ab}}^i\;,\\ %
  [X_a,X_5] ={}& \sfrac{3}{2\varrho} \, \eta^{1\ b}_{\ a}\, X_b +
\sfrac{2}{3}\, {f_{a5}}^b\, \dot{X}_b\;.
 \end{align}
\label{eqn:inst-eqns_HF}%
\end{subequations}
One can employ the following ansatz:
\begin{equation}
 X_a = \psi(r) \left(\exp(\xi \ \h^1) \exp(\theta \ \h^2) \right)_{a}^{\ b}
\hat{I}_b \; , \for \theta, \xi \in [0,2 \pi) \and X_5 = \chi(r) \hat{I}_5 \; , 
\label{eqn:ansatz_HF2}
\end{equation}
which, again, satisfies the equivariance condition. The insertion of
\eqref{eqn:ansatz_HF2} into \eqref{eqn:inst-eqns_HF} yields for the
$\mathfrak{su}(2)$-part
\begin{equation}
 \mathcal{N}_i = \psi^2 I_i\;,
\end{equation}
as the projection of $[X_a,X_b]$ onto $\mathfrak{su}(2)$ in $\mathfrak{su}(3)$
is independent of $\theta$ and $\xi$. Further, for the functions $\psi$ and 
$\chi$ one derives the set of equations
\begin{subequations}
 \begin{align}
  \dot{\chi} ={}& \frac{3}{\varrho^2}\,\psi^2\,(\cos^2\!\theta -
\sin^2\!\theta) \;,\\ %
  \chi ={}& \frac{2}{\varrho}\,\psi^2\,\cos\theta\,\sin\theta\;,\\%
  \dot{\psi}\,\cos\theta ={}& \frac{3}{2}\,\psi \left( 
\frac{1}{\varrho}\,\sin\theta  + \chi\,\cos\theta\right) \;,\\%
  \dot{\psi}\,\sin\theta ={}& -\frac{3}{2}\,\psi \left(
\frac{1}{\varrho}\,\cos\theta + \chi\,\sin\theta \right) \;.%
 \end{align}%
\label{eqn:reduced_inst-eqns_HF}%
\end{subequations}%
Note that the equations are independent of $\xi$. These equations are mutually
compatible only for $\theta{=}\frac{\pi}{4}$ or $\theta{=}\frac{3 \pi}{4}$.
For these values of $\theta$ the first two
equations yield $\dot{\psi} = \dot{\chi} = 0$ and the last two
equations coincide.
The system~\eqref{eqn:reduced_inst-eqns_HF} admits, besides the trivial
solution $(\psi,\chi)=(0,0) $, only the following solutions:
\begin{subequations}
 \begin{alignat}{3}
  \theta &=\frac{\pi}{4} & \;&: \qquad &  \psi &= \pm 1 \; , \quad \chi = +
\frac{1}{\varrho}\;,\\
  \theta &=\frac{3 \pi}{4} & \;&: \qquad &  \psi &=  \pm 1 \;, \quad \chi = -
\frac{1}{\varrho}\;.
 \end{alignat}
\end{subequations}
 Hence, we again have a whole family of solutions given by 
\begin{equation}
 A = \Gamma + \left(\exp(\xi \ \h^1) \exp(\theta \ \h^2) \right)_{a}^{\ b} \,
\hat{I}_b \otimes e_z^a \pm \frac{1}{\varrho} \, \hat{I}_5 \otimes e_z^5  \; ,
\for
\theta \in \{ \sfrac{\pi}{4},\sfrac{3 \pi}{4} \} \;, \; \xi\in[0,2\pi) \; .
\end{equation}
As the corresponding instantons on the cylinder over $M^5$ do neither depend on
the cone coordinate nor contain $\diff r$, they are actually lifts of instantons
on $M^5$, which live on the pull-back bundle of the SU$(3)$-bundle on the slices 
of the cylinder.

%% file: diagramm.tex
 \begin{center}
 \begin{tikzpicture}  
  \draw[fill=black!10](0,0.25)..controls(5,1) and
(7,0)..(10,0.5)--(10,2.25)..controls(7,1.5) and (5,3)..(0,2)--cycle;
  \node(P) at (2,2.05){$\mathcal{P}$};
  
  \draw[fill=black!10](0,3.2)..controls(3,2) and
(7,4)..(10,3.2)--(10,4.9)..controls(7,5) and (3,4.2)..(0,5)--cycle;
  \node(P') at (2,4.25){$\mathcal{P}'$};
  \draw[->](2,5.5)--(P') node[pos=0,above]{\footnotesize{defines instantons
via $\Om' \wedge \mathcal{F}=0$ and $\diff \Om' \wedge \mathcal{F}=0$}};
  
  \draw[fill=black!20](0,0.5)..controls(5,1.5) and
(7,0)..(10,1)--(10,2)..controls(7,1) and (5,2.5)..(0,1.5)--cycle;
  \node(Q) at (1,1.25) {$\mathcal{Q}$};
  
  \draw[fill=black!20](0,3.5)..controls(3,2) and
(7,4.5)..(10,3.5)--(10,4.5)..controls(7,5) and (3,3)..(0,4.5)--cycle;
  \node(Q') at (1,3.65) {$\mathcal{Q}'$};

  \draw[->](5.5,1.25)--(5.5,4) node[pos=.53,right]{$R_L$};
  
  
  \draw[-](2.5,1.3)..controls(3,1.5) and (3.5,0.8)..(4,1.3)
node[pos=.5,above]{$e$};
  \draw[-](2.5,3.3)..controls(3,3.5) and (3.5,3.2)..(4,3.3)
node[pos=.5,above]{$R_L \circ e$};
  
  \node(GP) at (2,-1){\footnotesize{$\Gamma^P$ $\mathfrak{su}(2)$-valued here}};
  
  \draw[->] (GP) to (Q);
  
  \draw[line width=2pt](0,0) rectangle (10,5);
  \draw[line width=2pt](0,-2)--(10,-2) node[pos=.5,below]{$M^5{\times_\phi}I$};
  \draw[->](5.5,-.2)--(5.5,-1.8) node[pos=.5,right]{$\pi$};
  \draw[<->](10.25,0)--(10.25,5)
node[pos=.5,right]{$F\big(T(M^5{\times_\phi}I)\big)$}; 
 \end{tikzpicture}
 \end{center}

%% file: conclusion.tex
\section{Conclusions}
\noindent
We investigated the geometry of cylinders, cones and sine-cones over
$5$-dimensional SU$(2)$-manifolds. On the resulting $6$-dimensional conical
SU$(3)$-manifolds we formulated generalized instanton equations and reduced
them to matrix equations via the ansatz~\eqref{eqn:ansatz_generic}. In
particular, we focused on nearly Kähler and half-flat SU$(3)$-manifolds,
whereas previous work~\cite{Bunk:2014kva} had dealt with the
Kähler-torsion (KT) and hyper-Kähler-torsion (HKT) cases.

In particular, we constructed a nearly Kähler 6-manifold as a sine-cone over an 
arbitrary Sasaki-Einstein 5-manifold 
by means of a rotation of the SU(2)-structures on the slices. Employing the ansatz~\eqref{eqn:ansatz_NK}, 
the instanton equation was reduced to the set~\eqref{eqn:matrix-eq_NK} of matrix equations, 
for which we found a family of non-trivial, but constant solutions.
All of these correspond to lifts of $M^5$-instantons to $C_s(M^5)$. 
In addition, in Subsection 4.2.2 we obtained an instanton solution on the
manifold $C_s(M^5)$ 
by the construction of its $\mathfrak{su}(3)$-valued canonical connection. 
We decomposed this connection $\Gamma_{\mathfrak{su}(3)}$ into another $\mathfrak{su}(2)$-valued instanton 
$\Gamma_{\mathfrak{su}(2)}$ plus an additional part resembling the ansatz used before. 
Using this decomposition and, again, carrying the reduction of the 
instanton equation out, 
we obtained a set of four equations for two functions which parametrize the ansatz. 
Its three solutions, for which the scalar functions take certain constant values, 
correspond to three instantons on the nearly Kähler sine-cone 
that cannot be constructed as lifts of instanton connections on $M^5$. 
As a by-product, we explicitly confirmed the nearly Kähler canonical connection 
to be an instanton. In addition, observing a correspondence between the 
solutions, we transferred the solutions of the two cases to new $r$-dependent 
instanton extensions of $\Gamma^P$ as well as $\Gamma_{\mathfrak{su}(2)}$. 
Remarkably, the extension found for $\Gamma^P$ does not seem to correspond to a 
lift of an instanton from $M^5$. 

Furthermore, we introduced a two-parameter family of half-flat structures on 
the cylinder over a generic Sasaki-Einstein 5-manifold. 
Again employing the ansatz~\eqref{eqn:ansatz_generic} on these cylindrical 
half-flat 6-manifolds, 
we were able to deduce the matrix equations~\eqref{eqn:inst-eqns_HF} on the two
local frames $e^{\hat{\mu}}$ and $e_z^{\hat{\mu}}$. Moreover, we provided
families of constant, but non-trivial solutions. In that case, the instantons
obtained this way do correspond to lifts of instantons on $M^5$.

It would be interesting to extend the methods presented here, i.e. the 
reduction of the instanton equation to matrix equations and the construction 
of higher-dimensional $G$-structure manifolds from lower-dimensional ones, to 
other scenarios that appear in string theory. For example, in M-theory desirable 
(internal) manifolds are $7$-dimensional and are endowed with a 
$G_2$-structure. Therefore, the study of certain SU$(3)$-structures seems to be 
promising, as one could hope to obtain interesting $G_2$-geometries as well as 
explicit instanton solutions via the procedures employed here.

Returning to the heterotic supergravity point of view, we expect that our 
solutions to the instanton equations can be lifted to full solutions of the 
heterotic equations of motions via the BPS equations~\eqref{eqn:BPS} and the 
Bianchi identity~\eqref{eqn:Bianci}. The gaugino equation~\eqref{eqn:gaugino} 
is already solved by the instanton solutions above. The remaining equations 
should be solvable in a manner similar 
to~\cite{Harland:2011zs,Gemmer:2012pp,Chatzistavrakidis:2012qb,Gemmer:2013ica}, 
which may look as follows: 
\begin{enumerate}
 \item The dilatino equation~\eqref{eqn:dilatino} may be solved by a 
suitable ansatz such as choosing the dilaton $\phi = \phi(\tau)$ and the 
$3$-form $H\propto \frac{\mathrm{d} \phi}{ \mathrm{d} \tau} \; P$ where $P$ is 
the canonical $3$-form on the Sasaki-Einstein $5$-manifold.
\item The gravitino equation~\eqref{eqn:gravitino} requires a spin connection 
with SU$(3)$-holonomy and torsion $H$. Therefore, one can take an ansatz 
similar to~\eqref{eqn:ansatz_generic} from which we know it to be an 
SU$(3)$-instanton. The remaining task is then to check the correct torsion for 
this connection. One choice might be the canonical connection 
$\Gamma_{\mathfrak{su}(3)}$ on the nearly Kähler sine-cone, whose torsion is by 
definition skew-symmetric, and we know $\Gamma_{\mathfrak{su}(3)}$ is 
an instanton.
  \item The theorem of Ivanov requires a connection $\nabla$ on $TM^6$ 
which is an instanton. Here, the instantons constructed in this paper provide a 
valuable choice, i.e. by an extension of the canonical connection. Then the 
connection $\nabla$, together with the gauge connection $\mathcal{A}$, needs to 
satisfy the Bianchi identity~\eqref{eqn:Bianci}. 
\end{enumerate}

Finally, one has to solve the differential equations that appear for the 
degrees of freedom in the different ansätze for $H $, $\nabla^+$, and $\nabla$. 
We hope to report on this process and embed our solutions into 
heterotic supergravity in the future.

\noindent
{\bf Acknowledgments}

\noindent
This work was partially supported by the Deutsche Forschungsgemeinschaft grant
LE 838/13.